\documentclass[12pt]{article}

\usepackage[a4paper,margin=2.5cm]{geometry}
\usepackage{graphicx}
\usepackage{amsmath}
\usepackage{amsfonts}
\usepackage{bm}
\usepackage{mathtools}
\usepackage[utf8]{inputenc}
\usepackage[english]{babel}
\usepackage[toc,page]{appendix}
\usepackage{authblk}
\usepackage{hyperref}
\usepackage{standalone}

\title{The moving bar problem: an electromechanical damped oscillator}
\author[1]{Carlos E. Alvarez\thanks{\href{mailto:carlose.alvarez@tec.mx}{carlose.alvarez@tec.mx}}}
\affil[1]{Tecnologico de Monterrey, School of Engineering and Sciences, 01389, Santa Fe, Mexico}
\date{\today}

\begin{document}
\maketitle

\begin{abstract}
The conducting bar sliding on rails through a uniform magnetic field is a
standard textbook illustration of Faraday's law, almost always solved
assuming the magnetic field produced by the induced current is negligible.
We extend this classic problem by retaining the self-induced field:
modelling the circuit as a rectangular loop of round wire of radius $d$, we
compute in closed form its geometry-dependent self-inductance $L(x,l)$ and
its gradient $dL/dx$ from the Biot--Savart law, including the flux inside
the wire and at the corners. The bar then obeys coupled
mechanical--electrical equations of motion containing, besides the familiar
braking force $-B_0lI$, the inductance-gradient force
$\tfrac{1}{2}I^2\,dL/dx$ familiar from electromagnetic launchers. In the
absence of resistance the total energy $\tfrac12Mv^2+\tfrac12LI^2$ is
exactly conserved; with resistance the system becomes an electromechanical
damped oscillator that, in an appropriate regime, maps onto a series
resistor--inductor--capacitor (RLC) circuit with equivalent capacitance
$C_{eq}=M/(l^2B_0^2)$, the bar's momentum playing the role of the
capacitor charge. Numerical integration of
the full equations confirms these analytic approximations in their
respective regimes and locates the crossover between over-damped and
under-damped behaviour. The full derivations, together with a runnable
Python implementation that reproduces every figure and re-runs every
numerical check reported here, are provided as supplementary material.
\end{abstract}


\section{Introduction}
\label{sec:intro}
A conducting bar of length $l$, sliding with speed $v$ along two rails,
closing a circuit of resistance $R$ in a uniform perpendicular magnetic
field $B_0$, is one of the canonical examples of electromagnetic induction,
appearing in virtually every introductory treatment of Faraday's law
\cite{feynman2011,griffiths2017}: the motional electromotive force (emf)
$\mathcal{E}=B_0lv$ drives a current $I=B_0lv/R$, whose interaction with the
external field brakes the bar, which decays exponentially to rest.

The problem has both a strong experimental and pedagogical tradition. It
has been realized directly on the bench, from a lecture demonstration with
modified model-railroad wheels \cite{sankovich1985} to an aluminium rod on
inclined conducting rails \cite{kwan2012} and a shorting bar dragged over an
array of neodymium magnets \cite{smith2020}, each confirming
$\mathcal{E}=B_0lv$ and its link to Newtonian mechanics and energy
conservation. Pedagogically, the problem has been used to separate motional
from transformer emf and expose the frame-dependence of that split
\cite{galili2006}, to clarify the role of the swept surface in applying the
flux rule \cite{zuza2012}, and to compare the flux-rule and Lorentz-force
routes to computing the emf \cite{scanlon1969}.

In all of these treatments the induced current is assumed small enough that
its own magnetic field is negligible \cite{feynman2011,griffiths2017}. When
self-inductance is included at all, it is usually as a separate, lumped
circuit element of fixed value \cite{saslow1987}, which already turns the
aperiodic decay into an oscillation. In the bare rails-and-bar circuit,
however, the inductance is not an added element: the current loop's
self-inductance is fixed by its own geometry, and since the bar's motion
changes that geometry, $L$ is a function of the bar's position $x$
(measured from the circuit's left end), $L=L(x,l)$. Closed-form
self-inductance expressions for rectangular circuits of round wire have
long been available \cite{rosa1908,grover2004}; what the moving-bar problem
adds is that the gradient $dL/dx$ acquires dynamical significance,
contributing both to the emf and to a force $\tfrac12I^2\,dL/dx$ on the bar
--- the same inductance-gradient thrust that propels electromagnetic
railguns \cite{mcnab2003}.

In this paper we solve the sliding-bar problem retaining the self-induced
field: we derive $L(x,l)$ and $dL/dx$ from the Biot--Savart law
(section~\ref{sec:selfind}), obtain the coupled equations of motion and
show that the total energy $\tfrac12Mv^2+\tfrac12LI^2$ is exactly conserved
without resistance (section~\ref{sec:energy}), identify a regime that maps
onto a damped RLC oscillator (section~\ref{sec:rlc}), and integrate the
full equations numerically to locate the crossover between the over-damped
and under-damped regimes (section~\ref{sec:simul}). Two supplementary
items accompany the paper: the \emph{supplementary derivations}, which
give in full the calculations outlined below, and the \emph{supplementary
code}, a set of documented Python modules and notebooks that reproduces
every figure shown here and re-runs every numerical check we report.

\section{Method}
\label{sec:method}
This section sets out the model and the analysis built on it: the standard
treatment of the sliding bar, its extension to include the self-induced
field, the approximate model that the extended equations reduce to in a
well-defined regime, and the numerical scheme and system of units used to
integrate them.

\subsection{The system and the standard treatment}
\label{sec:standard}
\begin{figure}[h]
  \centering
  \includegraphics[width=15cm]{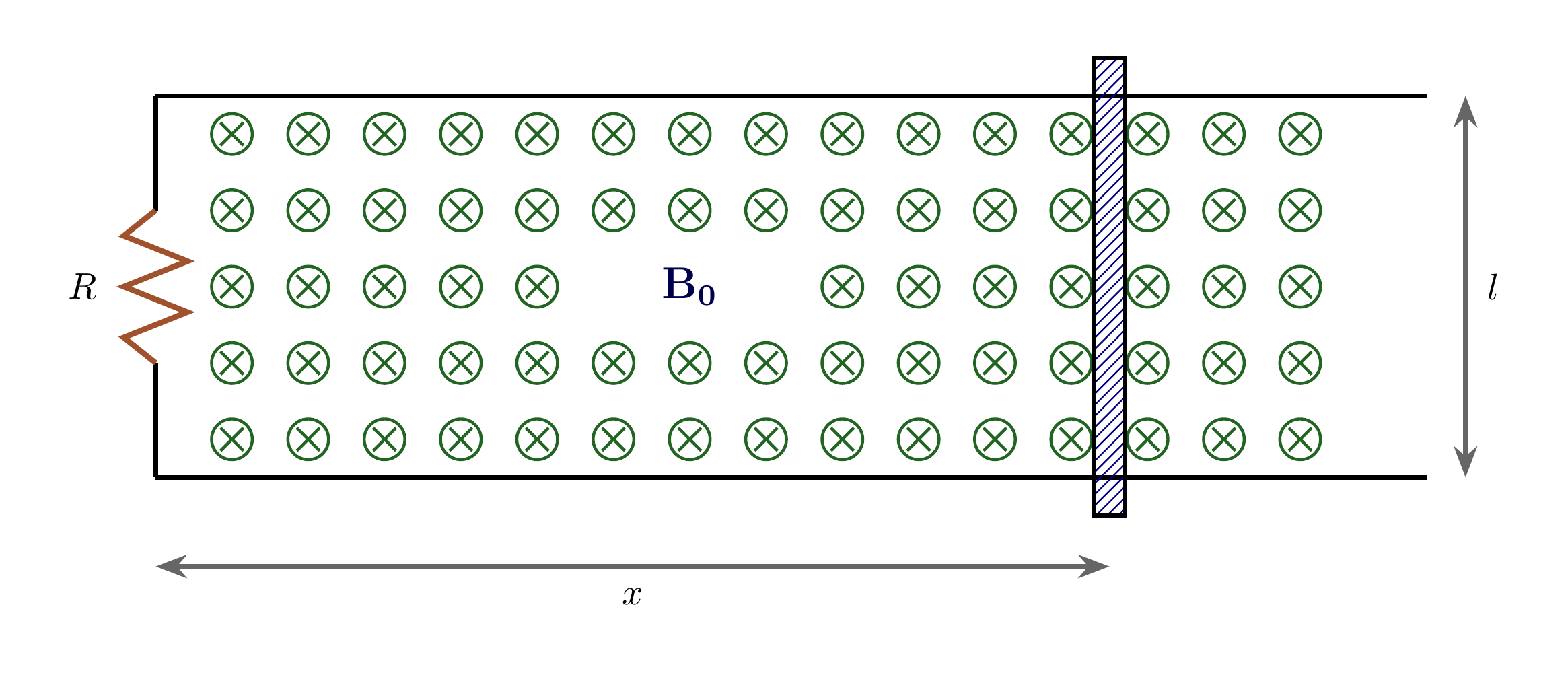}
  \caption{Geometry of the sliding-bar problem. A conducting bar closes an
    electric circuit of width $l$ and resistance $R$, so that when the bar
    moves, the area of the circuit, and with it the magnetic flux through
    the circuit, changes. A uniform magnetic field $B_0$ traverses the
    circuit perpendicularly, into the plane of the page, and the position
    $x$ of the bar is measured from the left end of the circuit, where the
    resistance sits.}
  \label{figcir}
\end{figure}

The system under study consists of a conducting bar of length $l$ that
closes a circuit of resistance $R$, in the presence of a perpendicular,
uniform magnetic field $B_0$ (figure \ref{figcir}). The bar slides over the
circuit, changing the enclosed area and generating a current $I$.

We first recall the standard treatment, in which the induced current is
assumed small enough that the field it produces is negligible
\cite{feynman2011}. The flux through the circuit is $\Phi=-B_0A=-B_0lx$,
with $x$ the bar's position measured from the circuit's left end. By
Faraday's law this drives an emf $\mathcal{E}=-d\Phi/dt=B_0lv$ and, with no
source in the circuit, a current $I=\mathcal{E}/R=B_0lv/R$. The same emf
can equivalently be obtained from the path integral of the Lorentz force
per unit charge along the bar, $\mathcal{E}=\oint(\bm{v}\times\bm{B}_0)\cdot
d\bm{l}$ \cite{scanlon1969}, which also gives the horizontal braking force
on the bar, $F_{mag,x}=-lB_0I$; only the bar's segment contributes, the
other three being at rest.

Using $I=B_0lv/R$, Newton's second law $M\dot{v}=-B_0^2l^2v/R$ integrates
to
\begin{align}
  v(t)&=v_0e^{-\alpha t},\label{vel}\\
  x(t)&=x_0+\frac{v_0}{\alpha}\left(1-e^{-\alpha t}\right),\label{pos0}
\end{align}
with
\begin{align}
  \alpha=\frac{B_0^2l^2}{RM}=\frac{I_0^2R}{2K_0},
  \label{invct}
\end{align}
where $I_0$ and $K_0$ are the bar's initial current and kinetic energy;
$\alpha$ is the inverse of the characteristic stopping time. The current
then follows as $I(t)=I_0e^{-\alpha t}$.

\subsection{Including the self-induced field}
\label{sec:selfind}
The previous treatment neglects any feedback from the current the bar's own
motion generates. We now retain it: the total field inside the loop is
\begin{align}
  \bm{B}(x,l;t)=\bm{B}_0+\bm{B}_{ind}(x,l;t),
  \label{totB}
\end{align}
with $\bm{B}_0=-B_0\hat{k}$ and $\bm{B}_{ind}$ the field of the current $I$
in the loop, computed from the Biot--Savart law. This assumes the current
is steady at all times, i.e.\ that changes in $I$ propagate instantaneously
around the loop --- the ``magnetic limit'' of the Galilean electrodynamics
of moving bodies, in which the displacement current is neglected and
Faraday's law keeps its usual form \cite{montigny2006,zangwill2013}.

We model the wire and bar as four cylinders of radius $d$ (non-magnetic,
like the surrounding medium, so $\mu_0$ is the only permeability that
appears), of length $x$ for the horizontal sides and $l$ for the vertical
ones, and compute $\bm{B}_{ind}$ segment by segment from the Biot--Savart
law, regularizing the interior of each wire at constant current density
\cite{jackson1999}; the explicit, piecewise expression is given in the
supplementary derivations. Because $\bm{B}_{ind}$ is linear in $I$, it factorizes as
\begin{align}
  \bm{B}_{ind}(p,h,t;x)=I(t)f_B(p,h;x)\hat{k},
  \label{totind}
\end{align}
with $f_B$ given explicitly in the supplementary derivations; we write $B_{ind}(p,h,t;x)$ simply as
$B_{ind}(p,h)$ except where the dependence on $t$ or $x$ matters
explicitly.

The magnetic flux is $\Phi=\Phi_0+\Phi_{ind}$, with
\begin{align}
  \Phi_0=-B_0lx
  \label{appfl}
\end{align}
the flux of the external field and $\Phi_{ind}=\int\bm{B}_{ind}\cdot
d\bm{a}$. At the four corners of the circuit two sides' cross sections
overlap and are regularized together (supplementary derivations); this corner
model is itself an approximation to the true current distribution of a
bent wire, and its relative contribution to flux and force vanishes only
logarithmically, rather than with the (small) corner area, as $d\to0$.

\subsubsection{Self-inductance}
\label{sec:L}
Integrating $\bm{B}_{ind}$ over the area enclosed by the circuit gives
\begin{align}
  \Phi_{ind}=I(t)L(x,l),
  \label{indfl}
\end{align}
where $L(x,l)$ is the self-inductance. Outside the wire every point of the
integration area is threaded by the full current $I$, so equating the flux
integral to $IL$ is exact; inside the wire this needs one refinement. At a
point a distance $r$ from a given side's own axis ($r<d$), only the
fraction $r^2/d^2$ of that side's current (uniform current density) is
enclosed by an infinitesimal sub-loop through the point, so only that
fraction of the local self-field is actually linked with the total current
$I$ appearing in $\Phi=IL$ (\ref{indfl}) --- the standard construction
giving an internal self-inductance of $\mu_0/8\pi$ per unit length, rather
than the naive (fully-linked) $\mu_0/4\pi$ obtained by integrating the
field itself with no such weight. This weighting is supplied by the
additive term $\Delta L(x,l)$, derived in the supplementary derivations,
giving the closed form
\begin{align}
  L(x,l)=\frac{\mu_0}{\pi}\Big[\mathcal{C}_1+\mathcal{C}_2+\mathcal{C}_3+\mathcal{C}_4+A_1+B_1+C_1+A_2+B_2+C_2\Big](x,l)+\Delta L(x,l),
  \label{Lxl}
\end{align}
with every term given explicitly in the supplementary derivations: the
calligraphic $\mathcal{C}_i$ collect the flux outside the wire, and the
roman $A_i$, $B_i$, $C_i$ the flux inside it and at the corners. Classical
closed-form
results for the self-inductance of rectangular loops of round wire
\cite{rosa1908,grover2004} are recovered in spirit by this calculation; the
flux-integral route followed here has the advantage of making the
dependence on the bar's position $x$, and hence the gradient $dL/dx$ that
drives the dynamics below, fully explicit.

The emf is $\mathcal{E}=-d\Phi/dt$, which with $\Phi=\Phi_0+IL(x,l)$ gives
\begin{align}
  \mathcal{E}=lB_0v-\frac{dL}{dx}vI-L\frac{dI}{dt},
  \label{emf2}
\end{align}
with the gradient
\begin{align}
  \frac{dL}{dx}=\frac{\mu_0}{\pi}\left[\zeta(x,l)-\zeta(x,d)+\frac{1}{\chi(x,l)}+\frac{x^2}{d^2}\left(\frac{1}{\chi(x,d)}-1\right)-1\right]+\frac{d\,\Delta L}{dx},
  \label{dL}
\end{align}
where $\zeta$ and $\chi$ are defined in appendix \ref{defs}, and $d\Delta
L/dx$, the derivative of the enclosed-current-fraction term, is given in
the supplementary derivations.

\subsubsection{Asymptotic behaviour}
\label{sec:asymp}
\begin{figure}[h]
  \centering
  \includegraphics[width=8.5cm]{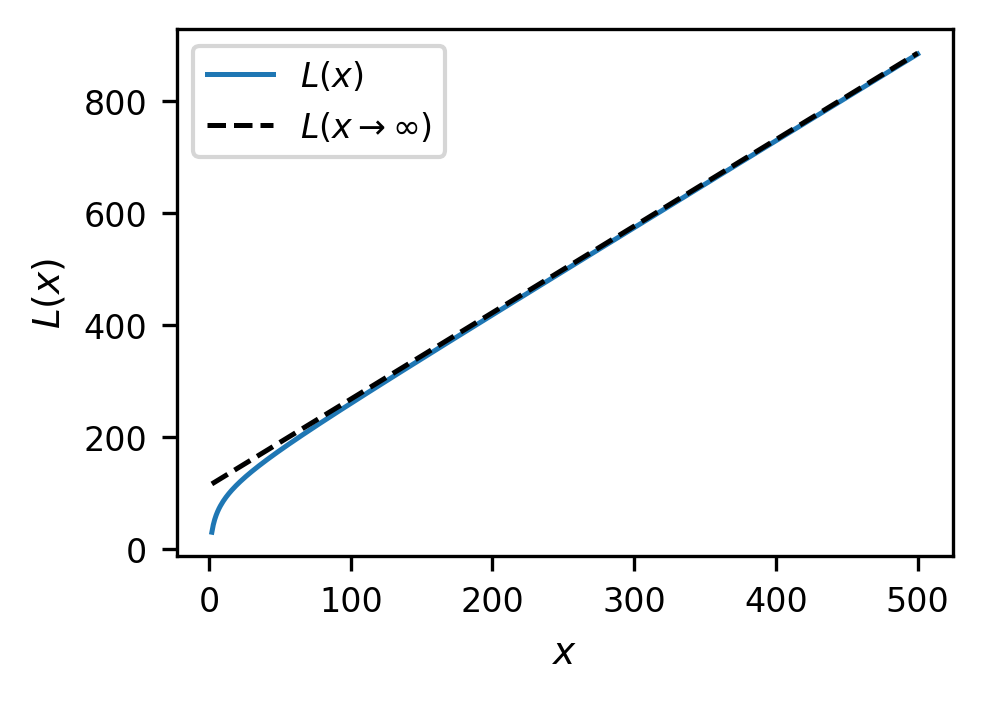}
  \caption{Self-inductance of the circuit as a function of the position of
    the bar. The solid line is the closed-form self-inductance $L(x)$ of
    the rectangular loop of round wire of radius $d$, exact for the model
    used here at every $x$; the dashed line is its large-$x$ asymptotic
    form, linear in $x$. The two differ by less than $3\%$ already at
    $x=100d$, so the linear form is an accurate stand-in once the bar is a
    hundred wire radii from the end of the circuit. Circuit width
    $l=100d$, and lengths are given in units of the wire radius $d$.}
  \label{Lafig}
\end{figure}

As $x\to\infty$, the self-inductance (\ref{Lxl}) takes the linear form
\begin{align}
  L(x,l)&\sim\frac{\mu_0}{\pi}\left(\ln\left|\frac{l}{d}\right|+\frac{1}{4}\right)x\nonumber\\
  &\hspace{0.3cm}+\frac{\mu_0}{3\pi d^2}\left[4d^3-3d^2l-l^3-3d^2l\,\zeta(l,d)+(l^2-2d^2)\sqrt{l^2+d^2}\right]+\frac{\mu_0}{15\pi}\left[\varphi(l)+4d\right],
  \label{Lasymp}
\end{align}
with $\varphi(v)=\left[2v^5+5d^2v^3-2\left(d^2+v^2\right)^{5/2}\right]/d^4$,
and correspondingly $dL/dx$
(\ref{dL}) tends to the constant
\begin{align}
  \frac{dL}{dx}\sim\frac{\mu_0}{\pi}\left(\ln\left|\frac{l}{d}\right|+\frac14\right).
  \label{dLasymp}
\end{align}
The enclosed-current-fraction term $\Delta L$ is what shifts this slope
from the naive $\ln|l/d|+\frac12$ to $\ln|l/d|+\frac14$ (supplementary
derivations). Figure \ref{Lafig} compares $L(x,l)$ with
(\ref{Lasymp}): the error is already below $3\%$ at $x=100d$.

\subsubsection{Magnetic force on the bar}
\label{sec:force}
The bar carries a free (mobile) linear charge density $\lambda_f$; a
charge moving along the bar has horizontal velocity $v$, shared with the
bar, and a vertical drift velocity $u$ related to the current by
$I=\lambda_f u$. The Lorentz force on these charges, integrated along the
bar and evaluated with the total field $\bm{B}=\bm{B}_0+\bm{B}_{ind}$ at
the bar's position $x$, has components
\begin{align}
  F_{mag,x}&=-lB_0I+\frac{1}{2}\frac{dL}{dx}I^2,\label{xforce}\\
  F_{mag,y}&=\lambda_f lvB_0-\frac{1}{2}\frac{dL}{dx}\lambda_f vI,\label{yforce}
\end{align}
where the induced-field contribution to both follows from evaluating
$\int_0^lB_{ind}(x,h)\,dh$ along the bar; because every $(x-p)$-dependent
term vanishes at $p=x$, this integral simplifies to the exact identity
\begin{align}
  \int_0^{l}B_{ind}(x,h)dh=\frac{1}{2}\frac{dL}{dx}I,
  \label{bind3}
\end{align}
(checked against direct numerical integration) between the field evaluated
on the bar's axis and the fully-linked part of $dL/dx$ (\ref{dL}).
(\ref{xforce}) and (\ref{yforce}) are therefore exact only to the extent
that the bar's own finite radius can be neglected in evaluating the
Lorentz force on it, i.e.\ that the field is uniform across the bar's
cross section and equal to its on-axis value --- even though that same
radius is properly accounted for in the flux-linkage calculation of
$L(x,l)$ itself.

We have checked the quality of this filamentary approximation directly, by
integrating the Lorentz force over the bar's actual cross section. The
bar's own field drops out of the force exactly, by symmetry, so the bar
exerts no net force on itself, as required. The strip average of the other
three segments' field, however, exceeds the on-axis value (\ref{bind3}) by
a universal offset of about $2.5\%$ of $dL/dx$ at $l/d=100$ --- comparable
to the uncertainty already introduced by the corner treatment of the field
model (supplementary derivations), and in the wrong direction to reconcile the
two. We therefore retain the filamentary force
(\ref{xforce})-(\ref{yforce}) for its closed form, and use the
energy-consistent $F=\frac12I^2\,dL/dx$, exact at constant current
regardless of the microscopic force distribution, in the equations of
motion and simulation code below, where it guarantees the exact energy
conservation of section \ref{sec:energy} in the perfect-conductor limit
(zero resistivity, $\rho=0$; resistivity $\rho$ is introduced below in
that section); the full check is reproduced in the supplementary
code.

The second term of (\ref{xforce}) is always positive, independent of the
current's direction: it is the same inductance-gradient thrust, always
repulsive between antiparallel-current segments, that at much larger
currents propels electromagnetic railguns \cite{mcnab2003}.

\subsubsection{Consistency between force and emf}
\label{sec:consistency}
As a check, the emf can also be obtained directly from Faraday's law, as
the path integral of the non-conservative field $\bm{v}\times\bm{B}$ along
the bar plus the contribution of the back-emf field $\bm{E}_{ind}$ induced
by the time-varying $\bm{B}_{ind}$ elsewhere in the loop,
\begin{align}
  \mathcal{E}=\oint(\bm{v}\times\bm{B})\cdot d\bm{l}+\oint\bm{E}_{ind}\cdot d\bm{l}.
\end{align}
Evaluating both terms (appendix \ref{sec:consistency-app}) reproduces
(\ref{emf2}) exactly. The calculation shows that the two halves of the
$\frac{dL}{dx}vI$ term in (\ref{emf2}) have distinct physical origins: one
half comes from the sliver of area swept by the moving bar, the other from
the change of the induced field throughout the rest of the loop as the bar
moves.

\subsubsection{Energy}
\label{sec:energy}
The power delivered by the circuit is $P_{circ}=I\mathcal{E}$, which, using
(\ref{emf2}), $\frac{dL}{dx}v=\frac{dL}{dt}$, and
$\frac{d}{dt}(\frac12LI^2)=LI\dot{I}+\frac12\dot{L}I^2$, becomes
\begin{align}
  P_{circ}=vB_0lI-\frac{1}{2}\frac{dL}{dt}I^2-\frac{d}{dt}\left(\frac{1}{2}LI^2\right).
  \label{pcirc}
\end{align}
In the absence of resistance, Kirchhoff's law sets $P_{circ}=0$. Adding the
mechanical power $P_{mech}=\frac{d}{dt}\left(\frac12Mv^2\right)=vF_{mag,x}=-vB_0lI+\frac12\dot{L}I^2$
(using (\ref{xforce})) to (\ref{pcirc}) then gives
\begin{align}
  \frac{d}{dt}\left(\frac{1}{2}Mv^2+\frac{1}{2}LI^2\right)=0,
  \label{cons1}
\end{align}
so that the energy
\begin{align}
  E=K+E_B=\frac{1}{2}Mv^2+\frac{1}{2}LI^2
  \label{econs1}
\end{align}
is exactly conserved, with
\begin{align}
  K&=\frac{1}{2}Mv^2,\label{kin}\\
  E_B&=\frac{1}{2}LI^2.\label{potB}
\end{align}

Because the magnetic force on each mobile charge is always perpendicular to
its velocity, it can do no work, and the external field supplies no energy
to the circuit \cite{mosca1974,griffiths2017}; its role is instead to
redirect energy. The vertical component of the force, $F_{mag,y}$
(\ref{yforce}), pushes charges along the bar and sustains the emf, while
the horizontal component $F_{mag,x}$ (\ref{xforce}) reacts back on the
bar. The actual work is done by whatever drives the bar against this
reaction --- here, the bar's own kinetic energy $K$ (\ref{kin}), drained as
it decelerates --- and the magnetic force merely channels that work into
the field energy $E_B$ (\ref{potB}), which, once the wire has a finite
resistivity, is ultimately dissipated as heat.

With resistivity $\rho$, the resistance of the circuit is
\begin{align}
  R=\frac{2(l+x)}{\pi d^2}\rho,
  \label{resist}
\end{align}
and energy is dissipated as heat at a rate $I^2R$, accumulating
\begin{align}
  Q=\int I^2R\,dt.
  \label{heat}
\end{align}

\subsubsection{Equations of motion}
\label{sec:eom}
Applying Kirchhoff's voltage law to the loop, using the emf (\ref{emf2}),
gives a differential equation for the current,
\begin{align}
  B_0lv-\left(R+\frac{dL}{dx}v\right)I-L\frac{dI}{dt}=0,
  \label{difeqI}
\end{align}
where the term $\frac{dL}{dx}v$ enters on the same footing as $R$: the
bar's motion acts on the current as a velocity-dependent ``motional
resistance'' (of either sign). The equations of motion are then
\begin{align}
  \frac{dx}{dt}&=v,\label{vel1}\\
  \frac{dv}{dt}&=\frac{F_{mag,x}}{M}=\frac{1}{M}\left[-lB_0I+\frac{1}{2}\frac{dL}{dx}I^2\right],\label{acc1}\\
  \frac{dI}{dt}&=\frac{1}{L}\left[lB_0v-RI-\frac{dL}{dx}vI\right].\label{curr1}
\end{align}

\subsection{Approximate model and RLC analogy}
\label{sec:rlc}
We now consider the regime in which $x$ is large enough for the asymptotic
form (\ref{Lasymp}) to hold, and the current is small enough that $I\ll
lB_0/(dL/dx)$; both conditions make the self-inductance terms in
(\ref{acc1}) and (\ref{curr1}) negligible next to the terms already present
in the standard treatment. The equations of motion then reduce to
\begin{align}
  \frac{dv}{dt}\approx-\frac{lB_0}{M}I,\qquad
  \frac{dI}{dt}\approx\frac{lB_0}{L}v-\frac{R}{L}I,
\end{align}
which combine into
\begin{align}
  \frac{d^2I}{dt^2}+\frac{R}{L}\frac{dI}{dt}+\frac{l^2B_0^2}{ML}I=0,
  \label{currsol}
\end{align}
the canonical damped-oscillator equation $\ddot{I}+2\gamma\dot{I}+\omega_0^2I=0$
with
\begin{align}
  \gamma=\frac{R}{2L},\qquad \omega_0=\frac{lB_0}{\sqrt{ML}}.
  \label{natfreq}
\end{align}
($\gamma$ plays here the role that $\alpha$ (\ref{invct}) played in the
no-self-inductance model, but is a distinct quantity: it damps the
current $I$ in this second-order equation, rather than the velocity $v$
in the first-order one of section~\ref{sec:standard}.) While $\gamma$ is
independent of $x$, $\omega_0$ decreases as $x^{-1/2}$, since $L$ and $R$
both grow linearly with $x$; we treat $\omega_0$ as constant, evaluated
at the bar's initial position $x_0$.

This is exactly the equation of a series RLC circuit,
$\ddot{I}+(R/L)\dot{I}+I/(LC)=0$, identifying an equivalent capacitance
$C_{eq}=M/(l^2B_0^2)$ and, from the energy stored in a capacitor $Q^2/2C$
compared with the bar's kinetic energy $Mv^2/2$, an equivalent charge
$q_{eq}=Mv/(lB_0)$ (not to be confused with the free charge $l\lambda_f$ in
the bar). The momentum of the bar thus plays the role of the charge
accumulated on the capacitor, and the bar's mass that of the capacitance.

The solution of (\ref{currsol}) is
\begin{align}
  I(t)=-I_{mx}e^{-\gamma t}\cos(\omega_dt+\phi),\qquad \omega_d=\sqrt{\omega_0^2-\gamma^2},
  \label{appi}
\end{align}
with $I_{mx}$ and $\phi$ fixed by $I(0)=I_0=-I_{mx}\cos\phi$. In this
regime the force (\ref{xforce}) reduces to
\begin{align}
  F_x(t)\approx-lB_0I=lB_0I_{mx}e^{-\gamma t}\cos(\omega_dt+\phi).
  \label{fapp}
\end{align}

\subsection{Algorithm}
\label{sec:algorithm}
Equations (\ref{vel1})-(\ref{curr1}) are integrated with a
velocity-Verlet, kick-drift-kick scheme \cite{verlet1967,swope1982,hlw2006}:
\begin{align}
  v_{i+1/2}&=v_i+\frac{\Delta t}{2M}\left[-B_0lI_i+\frac{1}{2}\left.\frac{dL}{dx}\right|_{x_i}I_i^2\right],\\
  x_{i+1}&=x_i+v_{i+1/2}\Delta t,\\
  I_{i+1}&=\frac{\left(1-\frac{\Delta t}{2}k\right)I_i+\Delta t\,\dfrac{B_0l}{L(x_m)}v_{i+1/2}}{1+\frac{\Delta t}{2}k},\qquad x_m=\frac{x_i+x_{i+1}}{2},\\
  v_{i+1}&=v_{i+1/2}+\frac{\Delta t}{2M}\left[-B_0lI_{i+1}+\frac{1}{2}\left.\frac{dL}{dx}\right|_{x_{i+1}}I_{i+1}^2\right],
\end{align}
with $k=\left[R(x_m)+\left.\dfrac{dL}{dx}\right|_{x_m}v_{i+1/2}\right]/L(x_m)$. The position/velocity
update is split into two half-kicks around a drift step; the circuit
equation (\ref{curr1}), linear in $I$ during the drift, is advanced in
closed form by the trapezoidal (Crank--Nicolson) rule, with $v$ frozen at
$v_{i+1/2}$ and $L$, $dL/dx$, $R$ evaluated at the drift midpoint $x_m$
rather than at either endpoint --- both choices are what keep the scheme
second order (see the supplementary derivations). A dedicated convergence
study, including the dissipationless limit $\rho=0$ where (\ref{econs1})
is exactly conserved by the continuous equations, confirms second-order
convergence ($|\varepsilon|\propto\Delta t^{2.0}$) and a worst-case
transient relative energy error of $\sim10^{-8}$ at the time step used
throughout ($\Delta t=2\times10^{-4}$).

\subsection{Units}
\label{sec:units}
A convenient laboratory scale uses grams, seconds, millimeters and
milliamperes, in which $\mu_0\approx1.26\times10^{-6}\
\text{g\,mm\,s}^{-2}\text{mA}^{-2}$. For the numerics we instead
adimensionalize using the bar's mass $M$, the wire radius $d$, and $\mu_0$,
together with the time constant
\begin{align}
  \tau_{ch}(l)=\frac{1}{\gamma}=\lim_{x\to\infty}\frac{2L(x)}{R(x)}=\frac{\mu_0d^2}{\rho}\left(\ln\left|\frac{l}{d}\right|+\frac{1}{4}\right),
  \label{chtime}
\end{align}
evaluated at $l=de^{3/4}$ and with the resistivity of copper,
$\rho_{Cu}\approx1.68\times10^{-2}\ \text{m}\Omega\cdot\text{mm}$ (chosen as
a representative conductor), to fix $\tau=\mu_0d^2/\rho_{Cu}$. The reduced
quantities are $t^*=t/\tau$, $l^*=l/d$, $m^*=m/M$,
$I^*=\tau\sqrt{\mu_0/d}\,I/\sqrt{M}$, $B^*=\tau\sqrt{d/\mu_0}\,B/\sqrt{M}$,
$\rho^*=\tau\rho/(\mu_0d^2)$ and $E^*=\tau^2E/(d^2M)$, in terms of which
$M$, $d$ and $\mu_0$ are all equal to $1$. Unless otherwise noted,
simulations use $M=10$g, $d=1$mm, $\tau=7.5\times10^{-5}$s; the asterisk is
dropped hereafter.

\section{Results}
\label{sec:simul}
\label{sec:results}
Simulations use $\Delta t=2\times10^{-4}$ and, unless stated otherwise,
$l=100$, $x_0=500$, $p_0=-1\times10^{-2}$ (reduced units).

\begin{figure}[h]
  \centering
  \includegraphics[width=8.5cm]{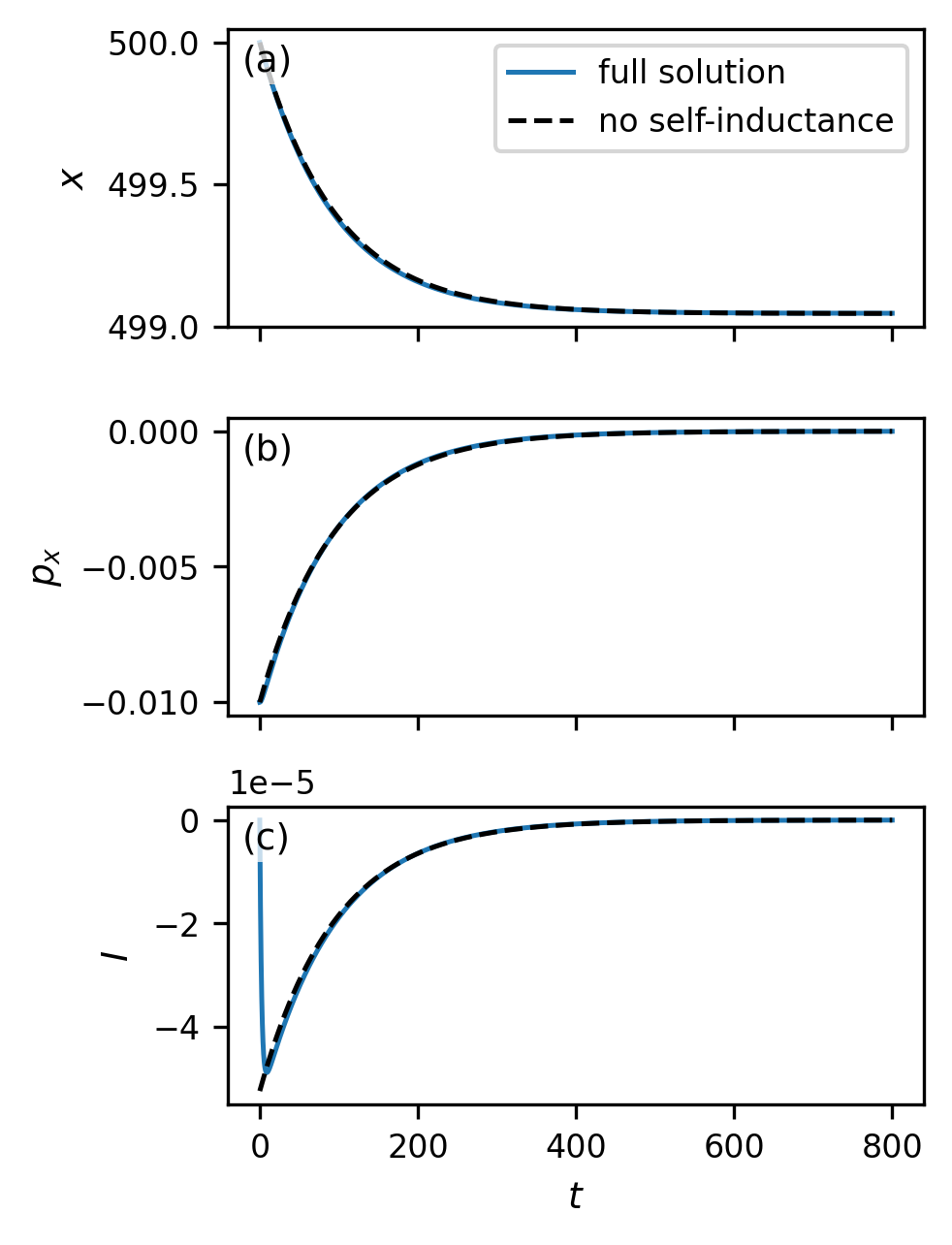}
  \caption{Over-damped motion of the bar: position $x$ (a), momentum $p_x$
    (b) and current $I$ (c) as a function of time, for a magnetic field
    $B_0=2\times10^{-2}$ and a resistivity $\rho=1.0$. Solid lines are the
    full solution, which retains the field induced by the current itself;
    dashed lines are the analytical solution of the standard treatment, in
    which the self-inductance of the circuit is neglected. The two agree
    closely, and the bar slows monotonically to rest without oscillating.
    Circuit width $l=100$, initial position $x_0=500$, initial momentum
    $p_0=-1\times10^{-2}$ and time step $\Delta t=2\times10^{-4}$, all in
    the reduced units used throughout.}
  \label{xpIvst_1}
\end{figure}

Figure \ref{xpIvst_1} shows $x$, $p_x$ and $I$ for copper resistivity
($\rho=1$) and $B_0=2\times10^{-2}$ ($\approx1$T): the standard
(no-self-inductance) solution (\ref{pos0}) agrees well with the full
solution, and the bar stops in about $0.05$s after moving $\sim1$mm at a
peak speed of $\sim13$cm/s.

\begin{figure}[h]
  \centering
  \includegraphics[width=8.5cm]{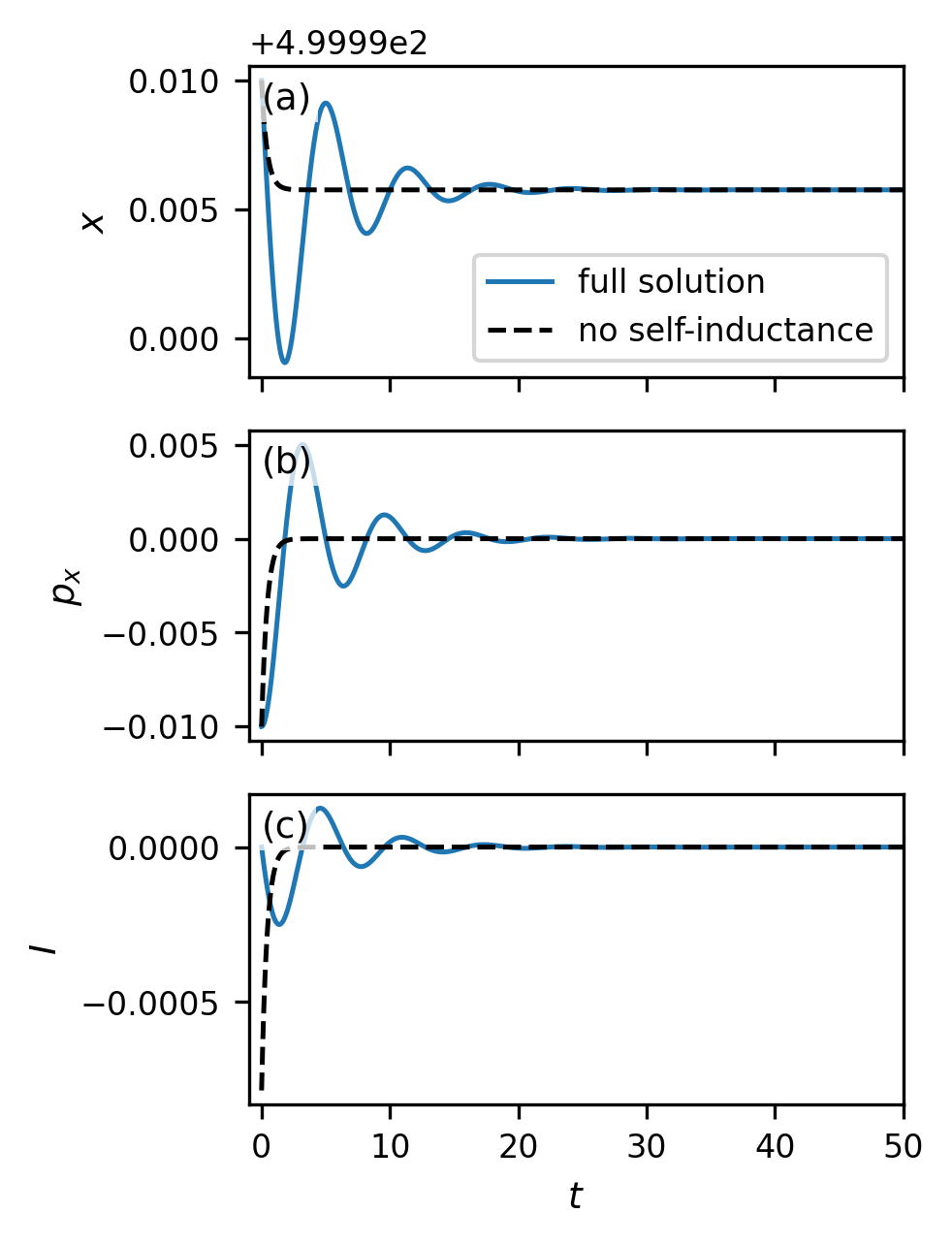}
  \caption{Under-damped motion of the bar: position $x$ (a), momentum
    $p_x$ (b) and current $I$ (c) as a function of time, for a magnetic
    field fifteen times larger than in the over-damped case,
    $B_0=3\times10^{-1}$, at the same resistivity $\rho=1.0$. Solid lines
    are the full solution, retaining the field induced by the current
    itself; dashed lines are the analytical solution of the standard
    treatment, which neglects the self-inductance of the circuit and no
    longer describes the dynamics: the bar now oscillates about an
    equilibrium position before stopping. Circuit width $l=100$, initial
    position $x_0=500$, initial momentum $p_0=-1\times10^{-2}$ and time
    step $\Delta t=2\times10^{-4}$, in reduced units.}
  \label{xpIvst_2}
\end{figure}

Figure \ref{xpIvst_2} shows the same quantities for a field fifteen times
larger. The system is now under-damped: the bar oscillates around an
equilibrium position before stopping, in about $0.002$s, with a small
initial amplitude ($\sim7\times10^{-3}$mm), and (\ref{pos0}) no longer
describes the dynamics.

\begin{figure}[h]
  \centering
  \includegraphics[width=8.5cm]{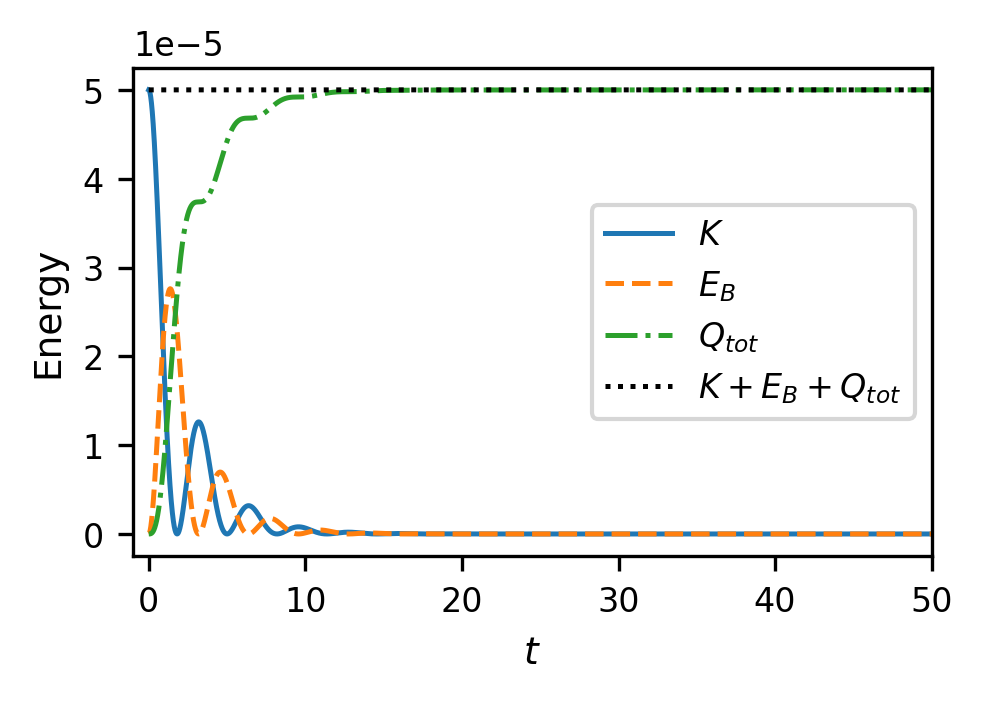}
  \caption{Energy bookkeeping in the under-damped regime, as a function of
    time, for a magnetic field $B_0=3\times10^{-1}$ and a resistivity
    $\rho=1.0$. The kinetic energy of the bar $K$ (solid), the magnetic
    energy stored in the circuit $E_B$ (dashed) and the heat dissipated in
    the resistance $Q_{tot}$ (dash-dotted) are exchanged as the bar
    oscillates and finally stops, while their sum (dotted) stays constant
    to a worst-case relative error of about $10^{-8}$. Circuit width
    $l=100$, initial position $x_0=500$, initial momentum
    $p_0=-1\times10^{-2}$ and time step $\Delta t=2\times10^{-4}$, in
    reduced units.}
  \label{econsfig}
\end{figure}

Figure \ref{econsfig} shows the kinetic (\ref{kin}), field (\ref{potB})
and dissipated (\ref{heat}) contributions to the energy for the
under-damped case: the energy (\ref{econs1}) decreases as the heat $Q$
accumulates, while the sum $K+E_B+Q$ stays conserved to a worst-case
transient relative error of $\sim10^{-8}$.

The values used above ($M=10$g, $d=1$mm, $\tau=7.5\times10^{-5}$s) are
typical of a physics laboratory, and at this scale producing even a barely
perceptible oscillation, with negligible bar-rail friction, requires a
large field ($\sim15$T) at an initial speed of $\sim13$cm/s. Scaling up to
$M=10$kg, $d=0.1$m and $\tau=0.75$s --- admittedly large but illustrating
the point --- the same initial speed instead produces oscillation
amplitudes of a couple of centimeters with $B_0\sim0.015$T.

\begin{figure}[h]
  \centering
  \includegraphics[width=8.5cm]{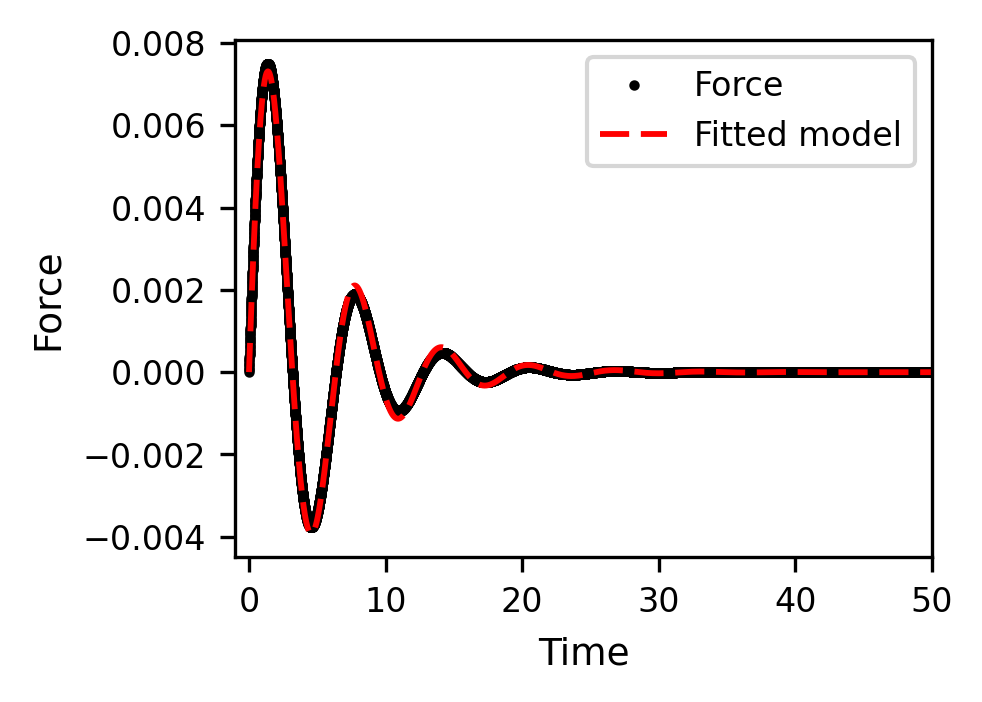}
  \caption{Total force on the bar as a function of time in the
    under-damped regime, for a magnetic field $B_0=3\times10^{-1}$ and a
    resistivity $\rho=1.0$. Points are the simulated force; the dashed
    line is the damped-oscillator model fitted to them, with the amplitude
    and the phase as the only free parameters, the decay rate and the
    frequency being fixed by the model itself. The fit reproduces the
    simulated force with a coefficient of determination above $0.996$.
    Circuit width $l=100$, initial position $x_0=500$, initial momentum
    $p_0=-1\times10^{-2}$ and time step $\Delta t=2\times10^{-4}$, in
    reduced units.}
  \label{fvst}
\end{figure}

Figure \ref{fvst} shows the force on the bar together with a fit of the
damped-oscillator model (\ref{fapp}) (using
\texttt{scipy.optimize.curve\_fit}), for $B_0=3\times10^{-1}$, $\rho=1.0$:
the fit achieves $\text{R}^2>0.996$ and RMSE$<10^{-4}$.

\begin{figure}[h]
  \centering
  \includegraphics[width=15cm]{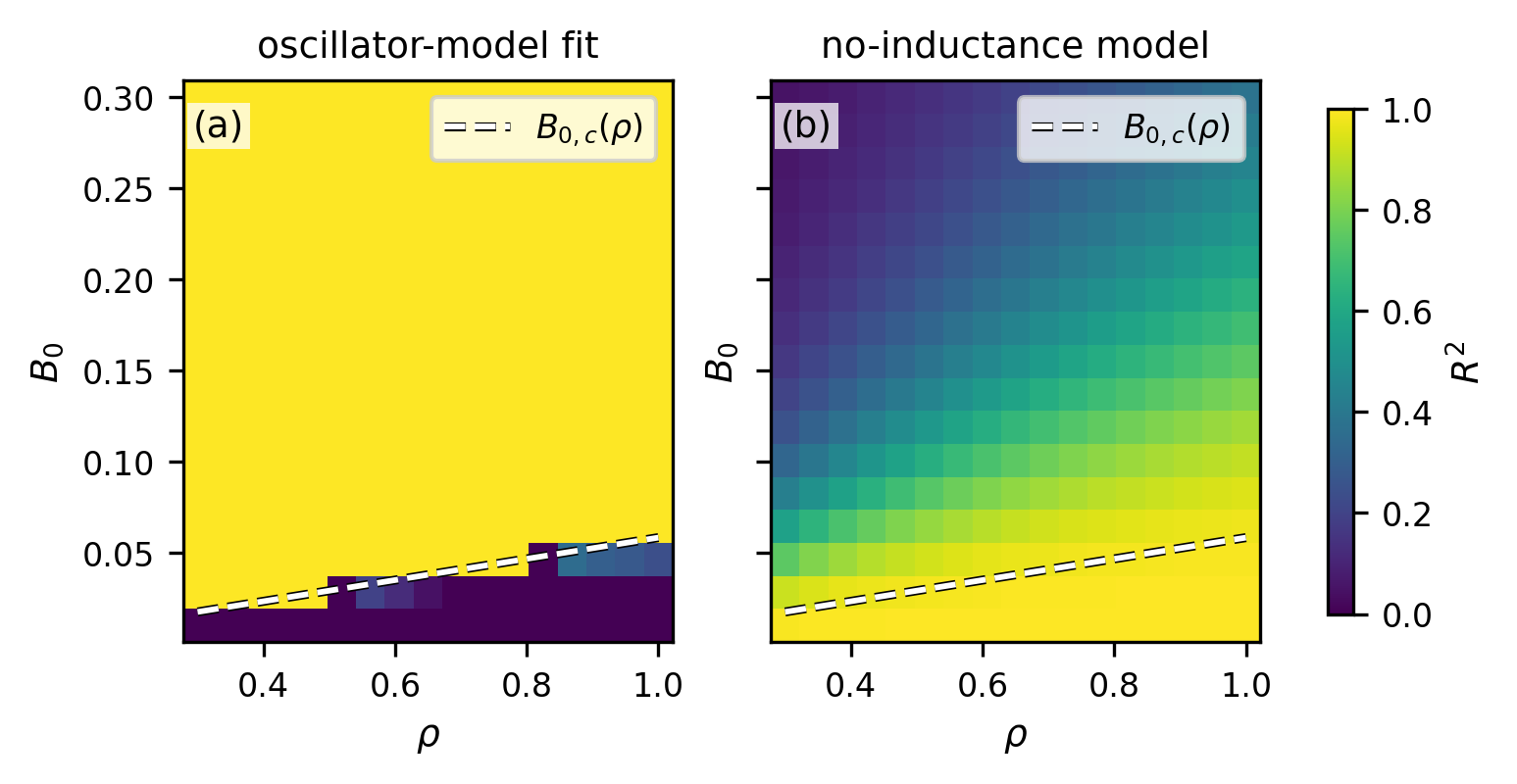}
  \caption{Coefficient of determination $\text{R}^2$ of nonlinear fits of
    the two approximate models to the simulated dynamics, as a function of
    the resistivity $\rho$ and the magnetic field $B_0$: (a) the
    damped-oscillator model, which is accurate above the critical field,
    and (b) the model that ignores the self-inductance of the circuit,
    which is accurate below it. The dashed white line in both panels marks
    the field $B_{0,c}(\rho)$ at which the system is critically damped;
    the quality of each fit drops sharply on crossing it, so the two
    models divide the parameter plane between them. Circuit width
    $l=100$, initial position $x_0=500$, initial momentum
    $p_0=-1\times10^{-2}$ and time step $\Delta t=2\times10^{-4}$, in
    reduced units.}
  \label{gfits}
\end{figure}

Figure \ref{gfits} shows the coefficient of determination for nonlinear
fits of the two approximate models, the no-self-inductance solution
(\ref{pos0}) and the damped-oscillator model (\ref{fapp}), as a function of
$B_0$ and $\rho$: the former is accurate in the over-damped regime
($B_0<B_{0,c}(\rho)$) and the latter in the under-damped regime
($B_0>B_{0,c}$), with
\begin{align}
  B_{0,c}(\rho)=\frac{M^{1/2}L^{1/2}}{l}\gamma(\rho)
  \label{bcrit}
\end{align}
the field at the critically damped point ($\omega_0=\gamma$); the
coefficient of determination drops sharply as $B_0$ crosses $B_{0,c}$ from
above.

\section{Discussion}
\label{sec:discussion}
We now place the closed-form self-inductance obtained above alongside the
classical results for rectangular loops, and then draw out what the whole
treatment offers a physics course.

\subsection{Comparison with previous expressions}
\label{sec:comparison}
Rosa and Grover calculated the self-inductance of a rectangle of
cylindrical wire of radius $d$, length $x$ and width $l$, under the
assumption that the circuit dimensions are much larger than the wire
radius (equation (24) of \cite{rosa1908}, and (58) of \cite{grover2004}):
\begin{align}
  L(x,l)&=\frac{\mu_0}{\pi}\left[x\ln\left|\frac{2x}{d}\right|+l\ln\left|\frac{2l}{d}\right|+2\sqrt{x^2+l^2}\right.\nonumber\\
    &\hspace{0.5cm}-x\sinh^{-1}\left(\frac{x}{l}\right)-l\sinh^{-1}\left(\frac{l}{x}\right)\left.-2(x+l)+\frac{1}{4}(x+l)\right].
  \label{Lrosa}
\end{align}
As $x\to\infty$, (\ref{Lrosa}) takes the asymptotic form
\begin{align}
  L(x,l)&\sim\frac{\mu_0}{\pi}\left\{x\left[\ln\left|\frac{l}{d}\right|+\frac{1}{4}\right]+l\left[\ln\left|\frac{2l}{d}\right|-\frac{7}{4}\right]\right\},
  \label{Lrosaasymp}
\end{align}
whose slope agrees exactly, for any $l/d$, with that of (\ref{Lasymp}) ---
a nontrivial cross-check of the enclosed-current-fraction term $\Delta L$,
without which the slope of (\ref{Lxl}) would exceed Rosa and Grover's by
$\mu_0/(4\pi)$.

\begin{figure}[h]
  \centering
  \includegraphics[width=8.5cm]{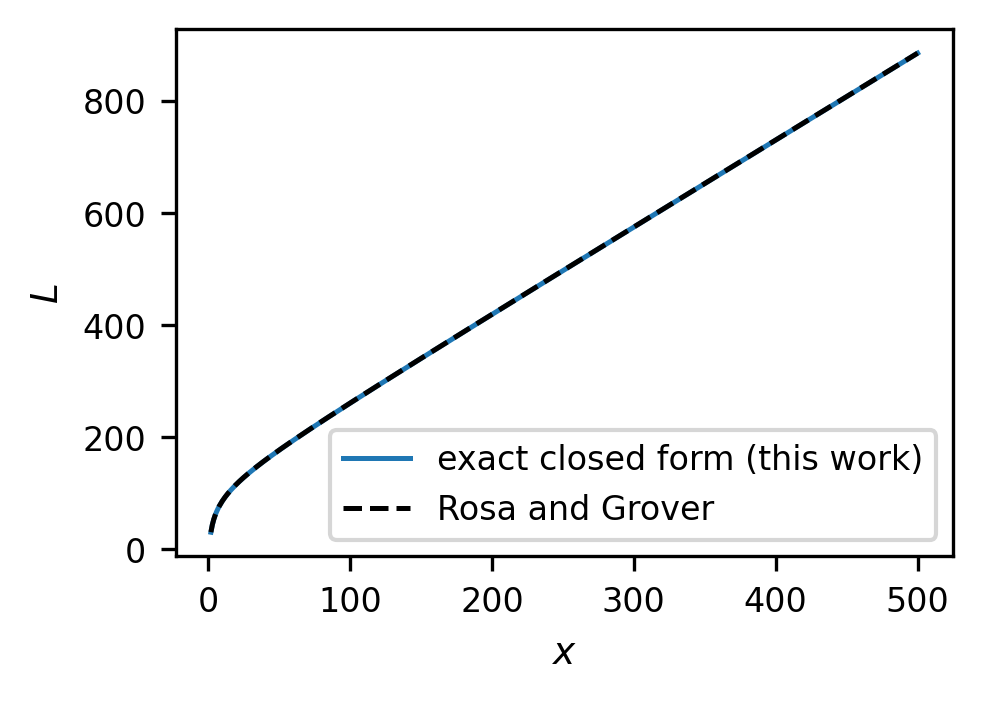}
  \caption{Self-inductance of the circuit as a function of the position of
    the bar: the closed form derived here (solid), exact for the model
    used at every $x$, and the classical expression for a rectangle of
    round wire of Rosa and Grover (dashed), which assumes that the
    dimensions of the circuit are much larger than the wire radius. The
    two curves are visually indistinguishable over the whole range shown,
    which certifies the classical formula down to bar positions of a few
    wire radii, where the assumption behind it no longer holds a priori.
    Circuit width $l=100d$, and lengths are given in units of the wire
    radius $d$.}
  \label{rosagrover}
\end{figure}

Unlike (\ref{Lrosa}), which assumes $x,l\gg d$, eq. (\ref{Lxl}) is exact
for the present model at all $x$, including $x\sim d$; figure
\ref{rosagrover} shows the two are visually indistinguishable over the
range plotted, with the ratio approaching $1$ as $x\to\infty$ and already
within $0.5\%$ for $x\geq100d$ at $l/d=100$. Even at $x=2d$ ($l/d=100$),
where the assumption underlying (\ref{Lrosa}) no longer holds a priori, it
remains within $2\%$ of (\ref{Lxl}) in $L$ and $0.1\%$ in $dL/dx$. Eq.
(\ref{Lxl}) therefore certifies (\ref{Lrosa}) down to small $x$, while
Rosa and Grover's formula independently tests (\ref{Lxl}) only
asymptotically, through the exact slope agreement above.

\subsection{Pedagogical significance}
\label{sec:pedagogy}
From a teaching perspective, the problem shows that a familiar system
contains, in an assumption usually made silently, a qualitatively
different dynamical regime; it exercises the Biot--Savart law in a
calculation that feeds directly into a dynamical problem; it gives a clean
example of energy bookkeeping in which the magnetic force does no work but
redirects energy between mechanical and magnetic reservoirs; and it
bridges mechanics and circuit theory through the RLC analogy. The
numerical side --- integrating three coupled first-order equations,
choosing units by dimensional analysis, and fitting analytic models to the
output --- is well within reach of an undergraduate computational physics
course, with each analytic approximation doubling as a validation test for
the student's code; the supplementary code package provides a ready
starting point for such a course module.

\section{Conclusion}
\label{sec:conclusions}
We have extended the sliding-bar problem to include the field generated by
the induced current itself. Modelling the circuit as a rectangular loop of
round wire, the Biot--Savart law yields a closed-form, position-dependent
self-inductance $L(x,l)$ whose gradient enters the dynamics twice: as an
additional emf term and as the always-repulsive force $\frac12I^2\,dL/dx$
on the bar. The resulting equations conserve $\frac12Mv^2+\frac12LI^2$
exactly for a perfect conductor, and reduce, in a well-controlled regime,
to a damped harmonic oscillator that maps term by term onto a series RLC
circuit, with the bar's momentum playing the role of the capacitor charge
and $C_{eq}=M/(l^2B_0^2)$ that of its capacitance. Numerical integration
confirms that the textbook and damped-oscillator approximations are each
accurate on their own side of the critical field $B_{0,c}$, cleanly
separating the over- and under-damped regimes.

Within the present scope, the natural frequency $\omega_0$ was treated as
constant, evaluated at the bar's initial position; for larger oscillation
amplitudes its position dependence should become evident, moving the
system into the regime of chirped oscillations, which we leave for future
work, along with a canonical (Lagrangian/Hamiltonian) formulation of the
same dynamics.

\bibliographystyle{unsrt}
\bibliography{references}

\begin{appendices}

\section{Definitions}\label{defs}
Define the following functions,
\begin{align}
  \chi(x,y)&=\frac{x}{\sqrt{x^2+y^2}},\label{chif}\\
  \zeta(x,y)&=\frac{1}{2}\ln\left|\frac{\sqrt{x^2+y^2}-x}{\sqrt{x^2+y^2}+x}\right|=-\tanh^{-1}\left(\chi(x,y)\right).\label{zetaf}
\end{align}
Note that the order of the parameters in (\ref{chif}) and (\ref{zetaf})
is non-commutative ($f(x,y)\neq f(y,x)$).

\section{Consistency between force and emf}\label{sec:consistency-app}
Using $B_{ind}(p,h;x)=I(t)f_B(p,h;x)$ (\ref{totind}), so that
$L(x)=\int_0^xdp\int_0^ldh\,f_B(p,h;x)$, Faraday's law applied to the bar
and to the back-emf field $\bm{E}_{ind}$
($\nabla\times\bm{E}_{ind}=-\partial\bm{B}_{ind}/\partial t$) gives
\begin{align}
  \mathcal{E}=\oint(\bm{v}\times\bm{B})\cdot d\bm{l}+\oint\bm{E}_{ind}\cdot d\bm{l}=vB_0l-\frac{1}{2}\frac{dL}{dx}vI-\int_0^xdp\int_0^ldh\left.\frac{\partial B_{ind}(p,h;x)}{\partial t}\right|_{p,h},
\end{align}
where the first term uses (\ref{yforce}) and only the bar contributes to
the path integral, the other three segments being at rest. At fixed
$(p,h)$, $f_B$ depends on time through $x(t)$, since Biot--Savart makes the
field everywhere in the loop depend on where the bar currently is, so
\begin{align}
  \left.\frac{\partial B_{ind}(p,h;x)}{\partial t}\right|_{p,h}&=\frac{dI}{dt}\,f_B(p,h;x)+Iv\,\frac{\partial f_B}{\partial x}(p,h;x).
\end{align}
Applying the Leibniz integral rule to $L(x)=\int_0^xdp\int_0^ldh\,f_B(p,h;x)$,
whose only $x$-dependent limit is the upper limit of $p$,
\begin{align}
  \frac{dL}{dx}&=\int_0^xdp\int_0^ldh\,\frac{\partial f_B}{\partial x}+\int_0^ldh\,f_B(x,h;x),
\end{align}
where the boundary term is exactly
$\frac{1}{I}\int_0^lB_{ind}(x,h)dh=\frac12\frac{dL}{dx}$ by (\ref{bind3});
hence the interior integral above is also $\frac12\frac{dL}{dx}$, and
\begin{align}
  -\int_0^xdp\int_0^ldh\left.\frac{\partial B_{ind}}{\partial t}\right|_{p,h}&=-L\frac{dI}{dt}-\frac12Iv\frac{dL}{dx}.
  \label{transf}
\end{align}
Combining (\ref{transf}) with the bar's contribution above gives
\begin{align}
  \mathcal{E}&=vB_0l-\frac12\frac{dL}{dx}vI-L\frac{dI}{dt}-\frac12Iv\frac{dL}{dx}=vB_0l-\frac{dL}{dx}vI-L\frac{dI}{dt},
\end{align}
which is exactly (\ref{emf2}). The two halves of the $\frac{dL}{dx}vI$
term thus have distinct origins: one half comes from the sliver of area
swept up at the bar, the other from the change of the induced field
throughout the rest of the loop as the bar moves.

\maketitle

\noindent
Closed-form Biot--Savart derivation of a sliding rectangular circuit's
self-inductance --- including in-wire and corner flux and
enclosed-current weighting --- plus the velocity-Verlet integrator's
second-order derivation.

\section{Setup and notation}\label{sec:setup}
These notes derive in closed form the self-inductance of a rectangular
circuit of round wire in which one side is free to slide, together with its
derivative with respect to that side's position. They are self-contained:
every symbol used below is defined either here or in section \ref{defs}.

The circuit is a rectangle of length $x$ and width $l$, made of four
cylindrical conductors of radius $d$ carrying a current $I$. The sliding
side (the bar) is the vertical segment at position $x$, the other three
sides being fixed. The conductors and the surrounding medium are taken to be
non-magnetic, so $\mu_0$ is the only permeability that appears. A point in
the plane of the circuit is labelled $(p,h)$, with $0\leq p\leq x$ measured
along the rails and $0\leq h\leq l$ across them, and $\hat{k}$ is the unit
normal to that plane.

The induced field is computed segment by segment from the Biot--Savart law,
regularizing the interior of each conductor at constant current density.
Being linear in the current, it factorizes as
\begin{align}
  \bm{B}_{ind}(p,h,t;x)=I(t)f_B(p,h;x)\hat{k},
  \label{totind}
\end{align}
and the flux it produces through the circuit defines the self-inductance
$L(x,l)$ through
\begin{align}
  \Phi_{ind}=I(t)L(x,l).
  \label{indfl}
\end{align}
The evaluation proceeds in two stages. The first stage is the fully-linked
flux integral
\begin{align}
  L_{fl}(x,l)=\int_0^xdp\int_0^ldh\,f_B(p,h;x),
  \label{Lfl}
\end{align}
assembled in sections \ref{sec:out}--\ref{sec:cor}. The second is an
additive term $\Delta L(x,l)$ (section \ref{sec:selfLcorr}), which accounts
for the fraction of each side's own current that is actually linked at
points lying inside that side, so that $L=L_{fl}+\Delta L$.

\section{Definitions and integrals}\label{defs}
Define the following functions,
\begin{align}
  \chi(x,y)&=\frac{x}{\sqrt{x^2+y^2}},\label{chif}\\
  \zeta(x,y)&=\frac{1}{2}\ln\left|\frac{\sqrt{x^2+y^2}-x}{\sqrt{x^2+y^2}+x}\right|=-\tanh^{-1}\left(\chi(x,y)\right).\label{zetaf}
\end{align}
Note that the order of the parameters in (\ref{chif}) and (\ref{zetaf})
is non-commutative ($f(x,y)\neq f(y,x)$), and we have the following
identities
\begin{align}
  \chi(x,y)&=\frac{x}{y}\chi(y,x),\\
  \frac{1}{\chi(x,y)}&=\chi(x,y)+\frac{y}{x}\chi(y,x).
\end{align}
We also define the symmetric combination
\begin{align}
  \eta(x,y)&=x\,\zeta(x,y)+y\,\zeta(y,x)=\eta(y,x),
  \label{etaf}
\end{align}
which appears below in (\ref{integ2}) and is used to simplify $A_1(x,l)$, $A_2(x,l)$ and $\mathcal{D}(x,l)$.

\begin{table}[h]
  \centering
  \begin{tabular}{|c|c|}
    \hline
    In terms of $x,y$ & In terms of $\chi,\zeta$\\
    \hline
    \hline
    $\frac{1}{2}\ln\left|\frac{\sqrt{x^2+y^2}-x}{\sqrt{x^2+y^2}+x}\right|$ & $\zeta(x,y)$\\
    \hline
    $\frac{x}{\sqrt{x^2+y^2}}$ & $\chi(x,y)$\\
    \hline
    $\sqrt{x^2+y^2}$ & $\frac{x}{\chi(x,y)}=\frac{y}{\chi(y,x)}$\\
    \hline
    $y\sqrt{x^2+y^2}$ & $\frac{xy}{\chi(x,y)}$\\
    \hline
    $\frac{xy}{\sqrt{x^2+y^2}}$ & $x\chi(y,x)=y\chi(x,y)$\\
    \hline
  \end{tabular}
\end{table}

We also present the following integral results,
\begin{align}
  \int\frac{dx}{y\ \chi(x,y)}&=\frac{1}{\chi(y,x)}+\zeta(y,x),
  \label{integ1}
\end{align}
\begin{align}
  \int\left(\frac{1}{\chi(y,x)}+\zeta(y,x)\right)dy&=y\zeta(y,x)+x\zeta(x,y)+\frac{2x}{\chi(x,y)},
  \label{integ2}
\end{align}
\begin{align}
  \int\frac{\chi(y,x)}{x}dx&=\zeta(y,x),
  \label{integ3}
\end{align}
\begin{align}
  \int\zeta(y,x)dy&=\frac{y}{\chi(y,x)}+y\zeta(y,x),
  \label{integ4}
\end{align}
\begin{align}
  \int x\chi(y,x)dx&=\frac{xy}{\chi(x,y)},
  \label{integ5}
\end{align}
\begin{align}
  \int \frac{xy}{\chi(x,y)}dy&=\frac{1}{3}\left(\frac{x}{\chi(x,y)}\right)^3,
  \label{integ6}
\end{align}
\begin{align}
  \int \frac{xy^3}{\chi(x,y)}dy&=\frac{1}{5}\left(\frac{x}{\chi(x,y)}\right)^5-\frac{x^2}{3}\left(\frac{x}{\chi(x,y)}\right)^3.
  \label{integ7}
\end{align}

\section{Field of a single wire segment}\label{sec:seg}
\begin{figure}[h]
  \centering
  \includegraphics[scale=0.3]{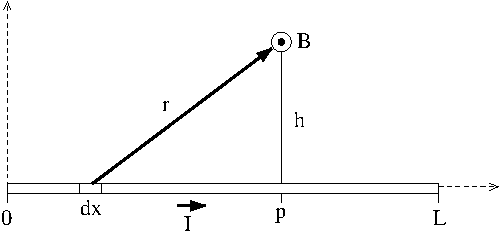}
  \caption{Magnetic field produced by a current element.}
  \label{biots}
\end{figure}

The Biot--Savart law for the induced magnetic field due to a steady
current moving in the $x$ direction outside the wire (figure \ref{biots})
reads
\begin{align}
  d\bm{B}'^{out}_{ind}&=\frac{\mu_0}{4\pi}\frac{I\bm{dx}\times\bm{r}}{r^3}=\frac{\mu_0 I}{4\pi}\frac{hdx}{\left((p-x)^2+h^2\right)^{3/2}}\hat{k},
  \label{bs0}
\end{align}
where $\mu_0$ is the magnetic permeability of free space; the wire, bar,
and surrounding medium are taken to be non-magnetic, so no distinct
material permeability appears.
The current's sign is fixed by the right-hand rule for a directed area
along $\hat{k}$ (counter-clockwise current for a positive area). The total
field at $(p,h)$ produced by a segment of length $w$ is
\begin{align}
  \bm{B}'^{out}_{ind}(p,h)&=\frac{\mu_0 Ih}{4\pi}\int_0^w\frac{dx}{\left((p-x)^2+h^2\right)^{3/2}}=\frac{\mu_0 I}{4\pi h}\left[\frac{w-p}{\sqrt{(w-p)^2+h^2}}+\frac{p}{\sqrt{p^2+h^2}}\right]\hat{k},
  \label{bs1}
\end{align}
using the substitution $u=h\tan\theta$ and
$\sin(\tan^{-1}(x/y))=x/\sqrt{y^2+x^2}$ in
\begin{align}
  \int_a^b\frac{du}{\left(u^2+h^2\right)^{3/2}}=\frac{1}{h^2}\left[\chi(b,h)-\chi(a,h)\right],
\end{align}
with $\chi(x,y)$ defined in section \ref{defs}. Equation (\ref{bs1}) is
valid outside the wire; inside, assuming the same permeability and a
constant current density $J=I/(\pi d^2)$, the enclosed current at distance
$h<d$ from the axis is $I'(h)=Ih^2/d^2$, and the field becomes
\begin{align}
  \bm{B}'^{in}_{ind}(p,h)&=\frac{\mu_0 Ih}{4\pi d^2}\left[\frac{w-p}{\sqrt{(w-p)^2+h^2}}+\frac{p}{\sqrt{p^2+h^2}}\right]\hat{k}.
  \label{bsi}
\end{align}

\section{Field outside the wire}\label{sec:out}
A calculation similar to (\ref{bs1}) and (\ref{bsi}) can be carried out
for the four sides of a rectangular circuit of length $x$ and width $l$,
and added to find the total induced field at the point $(p,h)$, taking
into account that if the direction of the current is positive in the
counterclock-wise sense, the limits of the integrals on each segment have
to be such that the induced field in the interior of the circuit points
in the positive $\hat{k}$ direction. The field outside of the wire
($d\leq p\leq x-d$ and $d\leq h\leq l-d$) is given by
\begin{align}
  \bm{B}^{out}_{ind}(p,h)&=\frac{\mu_0 I}{4\pi h}\left[\chi(x-p,h)+\chi(p,h)\right]\hat{k}\nonumber\\
  &+\frac{\mu_0 I}{4\pi (x-p)}\left[\chi(l-h,x-p)+\chi(h,x-p)\right]\hat{k}\nonumber\\
  &+\frac{\mu_0 I}{4\pi (l-h)}\left[\chi(p,l-h)+\chi(x-p,l-h)\right]\hat{k}\nonumber\\
  &+\frac{\mu_0 I}{4\pi p}\left[\chi(h,p)+\chi(l-h,p)\right]\hat{k},
  \label{bindp}
\end{align}
which can be written as
\begin{align}
  \bm{B}^{out}_{ind}(p,h)&=I\,f_B^{out}(p,h;x)\hat{k},
  \label{bind}
\end{align}
where
\begin{align}
  f_B^{out}(p,h;x)&=\frac{\mu_0}{4\pi}\left[\frac{1}{(x-p)\,\chi(h,x-p)}+\frac{1}{p\,\chi(h,p)}+\frac{1}{(x-p)\,\chi(l-h,x-p)}\right.\nonumber\\
    &\left.\hspace{0.5cm} +\frac{1}{p\,\chi(l-h,p)}\right].
  \label{bfunout}
\end{align}
Using (\ref{bind}), the flux due to the induced field outside the wire
can be written as
\begin{align}
  \Phi_{ind}^{out}&=I\int_{d}^{x-d}dp\int_{d}^{l-d}dh\ f_B^{out}(p,h;x),
\end{align}
where the first integral of (\ref{bfunout}) with respect to $h$ is
\begin{align}
  \int_{d}^{l-d}dh\ f_B^{out}(p,h;x)&=\frac{\mu_0}{2\pi}\left[g(x-p)+g(p)\right],
  \label{inth}
\end{align}
with
\begin{align}
  g(x)&=\zeta(x,l-d)-\zeta(x,d)+\frac{1}{\chi(x,l-d)}-\frac{1}{\chi(x,d)},
  \label{gfun}
\end{align}
and $\zeta(x,y)$ is defined in section \ref{defs}. The second integral,
with respect to $p$, gives
\begin{align}
  \int_{d}^{x-d}dp\int_{d}^{l-d}dh\ f_B^{out}(p,h;x)&=\frac{\mu_0}{\pi}\left[\mathcal{C}_1(x,l)+\mathcal{C}_2(x,l)+\mathcal{C}_3(x,l)+\mathcal{C}_4(x,l)\right]
  \label{cexpr}
\end{align}
with,
\begin{align}
  \mathcal{C}_1(x,l)&=(x-d)\left[\zeta(x-d,l-d)-\zeta(x-d,d)\right],\\
  \mathcal{C}_2(x,l)&=(l-d)\left[\zeta(l-d,x-d)-\zeta(l-d,d)\right],\\
  \mathcal{C}_3(x,l)&=-d\left[\zeta(d,l-d)+\zeta(d,x-d)-2\zeta(d,d)\right],\\
  \mathcal{C}_4(x,l)&=2\left[\frac{x-d}{\chi(x-d,l-d)}+\frac{d}{\chi(d,d)}-\frac{x-d}{\chi(x-d,d)}-\frac{d}{\chi(d,l-d)}\right].
\end{align}

\section{Field inside the wire}\label{sec:in}
The total induced field inside each circuit segment combines (\ref{bsi})
for that segment's own contribution with (\ref{bs1}) from the other three
segments. For the segment to the right (the bar), the field inside is
\begin{align}
\bm{B}_{ind}^{in,r}(p,h)&=\frac{\mu_0 I}{4\pi h}\left[\chi(x-p,h)+\chi(p,h)\right]\hat{k}\nonumber\\
&+\frac{\mu_0 I}{4\pi d^2}\left[(l-h)\,\chi(x-p,l-h)+h\,\chi(x-p,h)\right]\hat{k}\nonumber\\
&+\frac{\mu_0 I}{4\pi (l-h)}\left[\chi(p,l-h)+\chi(x-p,l-h)\right]\hat{k}\nonumber\\
&+\frac{\mu_0 I}{4\pi p}\left[\chi(h,p)+\chi(l-h,p)\right]\hat{k},
\end{align}
which, after reorganizing terms gives
\begin{align}
  B_{ind}^{in,rt}(p,h)&=I\,f_B^{in,rt}(p,h;x),
  \label{binrt}
\end{align}
where
\begin{align}
  f_B^{in,rt}(p,h;x)&=\frac{\mu_0}{4\pi}\left[\frac{1}{p\,\chi(h,p)}+\frac{1}{p\,\chi(l-h,p)}\right.\nonumber\\
    &\hspace{1cm}+\frac{\chi(x-p,h)}{h}+\frac{\chi(x-p,l-h)}{l-h}\nonumber\\
    &\left.\hspace{1cm}+\frac{(l-h)\,\chi(x-p,l-h)}{d^2}+\frac{h\,\chi(x-p,h)}{d^2}\right].
  \label{fbinrt}
\end{align}
The terms for the other three segments are obtained in a similar way.
\begin{align}
  B_{ind}^{in,lt}(p,h)&=I\,f_B^{in,lt}(p,h;x),
  \label{binlt}
\end{align}
where
\begin{align}
  f_B^{in,lt}(p,h;x)&=\frac{\mu_0}{4\pi}\left[\frac{1}{(x-p)\,\chi(h,x-p)}+\frac{1}{(x-p)\,\chi(l-h,x-p)}\right.\nonumber\\
    &\hspace{1cm}+\frac{\chi(p,h)}{h}+\frac{\chi(p,l-h)}{l-h}\nonumber\\
    &\left.\hspace{1cm}+\frac{(l-h)\,\chi(p,l-h)}{d^2}+\frac{h\,\chi(p,h)}{d^2}\right].
  \label{fbinlt}
\end{align}
\begin{align}
B_{ind}^{in,up}(p,h)&=I\,f_B^{in,up}(p,h;x),
  \label{binup}
\end{align}
where
\begin{align}
  f_B^{in,up}(p,h;x)&=\frac{\mu_0}{4\pi}\left[\frac{1}{p\,\chi(h,p)}+\frac{1}{(x-p)\,\chi(h,x-p)}\right.\nonumber\\
    &\hspace{1cm}+\frac{\chi(l-h,x-p)}{x-p}+\frac{\chi(l-h,p)}{p}\nonumber\\
    &\left.\hspace{1cm}+\frac{p\,\chi(l-h,p)}{d^2}+\frac{(x-p)\,\chi(l-h,x-p)}{d^2}\right].
  \label{fbinup}
\end{align}
\begin{align}
B_{ind}^{in,lo}(p,h)&=I\,f_B^{in,lo}(p,h;x),
  \label{binlo}
\end{align}
where
\begin{align}
  f_B^{in,lo}(p,h;x)&=\frac{\mu_0}{4\pi}\left[\frac{1}{p\,\chi(l-h,p)}+\frac{1}{(x-p)\,\chi(l-h,x-p)}\right.\nonumber\\
    &\hspace{1cm}+\frac{\chi(h,x-p)}{x-p}+\frac{\chi(h,p)}{p}\nonumber\\
    &\left.\hspace{1cm}+\frac{p\,\chi(h,p)}{d^2}+\frac{(x-p)\,\chi(h,x-p)}{d^2}\right].
  \label{fbinlo}
\end{align}

\section{Field at the corners}\label{sec:cor}
Each of $B_{ind}^{out}$, $B_{ind}^{in,rt}$, $B_{ind}^{in,lt}$,
$B_{ind}^{in,up}$ and $B_{ind}^{in,lo}$ is a sum of four contributions,
one from each side of the circuit, and only the contribution of the
side(s) whose cross section contains the point $(p,h)$ is replaced by its
regularized, uniform-current-density form (\ref{bsi}); the other sides
are always kept in their ordinary thin-wire form (\ref{bs1}). Near a
corner, the point $(p,h)$ lies simultaneously inside the cross section of
a vertical side (bar or left wire, $x-d\leq p\leq x$ or $0\leq p\leq d$)
and a horizontal side (bottom or top rail, $0\leq h\leq d$ or
$l-d\leq h\leq l$). Using the thin-wire form for either of these two
close sides diverges as the point approaches it, as can be seen for
example by letting $h\rightarrow0$ in the bottom-rail term of
$f_B^{in,rt}$ (\ref{fbinrt}). We therefore regularize \emph{both} nearby
sides at once, replacing their thin-wire terms by their (\ref{bsi})-type
forms, while the two far sides keep their thin-wire form as before. This
gives, for the bottom-right corner ($x-d\leq p\leq x$, $0\leq h\leq d$),
\begin{align}
  B_{ind}^{lo,rt}(p,h)&=I\,f_B^{lo,rt}(p,h;x),
  \label{binlort}
\end{align}
for the top-right corner ($x-d\leq p\leq x$, $l-d\leq h\leq l$),
\begin{align}
  B_{ind}^{up,rt}(p,h)&=I\,f_B^{up,rt}(p,h;x),
  \label{binuprt}
\end{align}
for the bottom-left corner ($0\leq p\leq d$, $0\leq h\leq d$),
\begin{align}
  B_{ind}^{lo,lt}(p,h)&=I\,f_B^{lo,lt}(p,h;x),
  \label{binlolt}
\end{align}
and for the top-left corner ($0\leq p\leq d$, $l-d\leq h\leq l$),
\begin{align}
  B_{ind}^{up,lt}(p,h)&=I\,f_B^{up,lt}(p,h;x),
  \label{binuplt}
\end{align}
with
\begin{align}
  f_B^{lo,rt}(p,h;x)&=\frac{\mu_0}{4\pi d^2}\left[(l-h)\,\chi(x-p,l-h)+2h\,\chi(x-p,h)+p\,\chi(h,p)\right]\nonumber\\
    &+\frac{\mu_0}{4\pi (l-h)}\left[\chi(p,l-h)+\chi(x-p,l-h)\right]+\frac{\mu_0}{4\pi p}\left[\chi(h,p)+\chi(l-h,p)\right],
  \label{fblort}
\end{align}
\begin{align}
  f_B^{up,rt}(p,h;x)&=\frac{\mu_0}{4\pi d^2}\left[2(l-h)\,\chi(x-p,l-h)+h\,\chi(x-p,h)+p\,\chi(l-h,p)\right]\nonumber\\
    &+\frac{\mu_0}{4\pi h}\left[\chi(x-p,h)+\chi(p,h)\right]+\frac{\mu_0}{4\pi p}\left[\chi(h,p)+\chi(l-h,p)\right],
  \label{fbuprt}
\end{align}
\begin{align}
  f_B^{lo,lt}(p,h;x)&=\frac{\mu_0}{4\pi d^2}\left[(l-h)\,\chi(p,l-h)+2p\,\chi(h,p)+(x-p)\,\chi(h,x-p)\right]\nonumber\\
    &+\frac{\mu_0}{4\pi (l-h)}\left[\chi(p,l-h)+\chi(x-p,l-h)\right]+\frac{\mu_0}{4\pi (x-p)}\left[\chi(l-h,x-p)+\chi(h,x-p)\right],
  \label{fblolt}
\end{align}
\begin{align}
  f_B^{up,lt}(p,h;x)&=\frac{\mu_0}{4\pi d^2}\left[2p\,\chi(l-h,p)+h\,\chi(p,h)+(x-p)\,\chi(l-h,x-p)\right]\nonumber\\
    &+\frac{\mu_0}{4\pi h}\left[\chi(x-p,h)+\chi(p,h)\right]+\frac{\mu_0}{4\pi (x-p)}\left[\chi(l-h,x-p)+\chi(h,x-p)\right].
  \label{fbuplt}
\end{align}
In each case the two regularized terms share a common piece (e.g.\
$h\,\chi(x-p,h)=(x-p)\,\chi(h,x-p)$ for (\ref{fblort})), which is why it
appears with a factor of 2. All four expressions are finite at the corner
point itself and match continuously with $B_{ind}^{in,rt}$,
$B_{ind}^{in,lt}$, $B_{ind}^{in,up}$ and $B_{ind}^{in,lo}$ across the
boundaries of their respective $d\times d$ domains, as verified
numerically.

Collecting the field in all nine regions defined above, the total
induced field can be written throughout the circuit as
$\bm{B}_{ind}(p,h,t;x)=I(t)f_B(p,h;x)\hat{k}$ (\ref{totind}), with
\begin{align}
  f_B(p,h;x)&=\begin{cases}
  f_B^{out}(p,h;x), & d<h<l-d\text{ and }d<p<x-d\\
  f_B^{in,lo}(p,h;x), & 0\leq h\leq d\text{ and }d<p<x-d\\
  f_B^{in,up}(p,h;x), & l-d\leq h\leq l\text{ and }d<p<x-d\\
  f_B^{in,lt}(p,h;x), & d<h<l-d\text{ and }0\leq p\leq d\\
  f_B^{in,rt}(p,h;x), & d<h<l-d\text{ and }x-d\leq p\leq x\\
  f_B^{lo,lt}(p,h;x), & 0\leq h\leq d\text{ and }0\leq p\leq d\\
  f_B^{up,lt}(p,h;x), & l-d\leq h\leq l\text{ and }0\leq p\leq d\\
  f_B^{lo,rt}(p,h;x), & 0\leq h\leq d\text{ and }x-d\leq p\leq x\\
  f_B^{up,rt}(p,h;x), & l-d\leq h\leq l\text{ and }x-d\leq p\leq x\\
  \end{cases},
  \label{bfun}
\end{align}
given explicitly by (\ref{bfunout}), (\ref{fbinrt})-(\ref{fbinlo}), and
(\ref{fblort})-(\ref{fbuplt}).

This construction is itself an approximation: the actual current
distribution around a real bent wire of finite radius, right at the
corner, can depart from the uniform circular cross section assumed here,
and a fully rigorous treatment would need to resolve that geometry. We do
not pursue that here. It is also worth noting that the corner
contribution to the flux and force does not vanish relative to the rest
of the self-inductance calculation as fast as its small ($d\times d$)
area might suggest: because the regularized terms scale as $1/d^2$ over a
region of size $d$, the corner integral tends to a constant as
$d\rightarrow0$ at fixed $x,l$, while the rest of the calculation grows
logarithmically; their ratio therefore vanishes only logarithmically
slowly, and remains at the level of a few percent even for fairly small
$d/l$.

The magnetic flux inside the wire is
\begin{align}
  \Phi_{ind}^{in}&=\int_{x-d}^{x}dp\int_{d}^{l-d}dh\ B_{ind}^{in,rt}(p,h)\nonumber\\
  &\hspace{0.5cm}+\int_{0}^{d}dp\int_{d}^{l-d}dh\ B_{ind}^{in,lt}(p,h)\nonumber\\
  &\hspace{0.5cm}+\int_{0}^{d}dh\int_{d}^{x-d}dp\ B_{ind}^{in,lo}(p,h)\nonumber\\
  &\hspace{0.5cm}+\int_{l-d}^{l}dh\int_{d}^{x-d}dp\ B_{ind}^{in,up}(p,h).
  \label{bflin0}
\end{align}
The flux through the four corners, using (\ref{binlort})-(\ref{binuplt}),
is
\begin{align}
  \Phi_{ind}^{cor}&=\int_{0}^{d}dp\int_{0}^{d}dh\ B_{ind}^{lo,lt}(p,h)\nonumber\\
  &\hspace{0.5cm}+\int_{0}^{d}dp\int_{l-d}^{l}dh\ B_{ind}^{up,lt}(p,h)\nonumber\\
  &\hspace{0.5cm}+\int_{x-d}^{x}dp\int_{0}^{d}dh\ B_{ind}^{lo,rt}(p,h)\nonumber\\
  &\hspace{0.5cm}+\int_{x-d}^{x}dp\int_{l-d}^{l}dh\ B_{ind}^{up,rt}(p,h).
  \label{bfcor0}
\end{align}

We use integral results (\ref{integ1}) to (\ref{integ6}). After
performing the integration in (\ref{bflin0}) and (\ref{bfcor0}), and
folding the corner contribution into $A_1$ and $A_2$ as described below,
we obtain
\begin{align}
  \Phi_{ind}^{in}+\Phi_{ind}^{cor}&=\frac{\mu_0 I}{\pi}\left[A_1(x,l)+B_1(l)+C_1(l)+A_2(x,l)+B_2(x)+C_2(x)\right],
  \label{philin}
\end{align}
where
\begin{align}
  A_1(x,l)&=\eta(x-d,d)-\eta(x-d,l-d)+\eta(x,l-d)-\eta(x,d)\nonumber\\
  &\hspace{0.3cm}+2\left[\sqrt{x^2+(l-d)^2}+\sqrt{(x-d)^2+d^2}-\sqrt{(x-d)^2+(l-d)^2}-\sqrt{x^2+d^2}\right]+\frac{1}{2}\mathcal{D}(x,l),
\end{align}
\begin{align}
  B_1(l)&=d\left[\zeta(d,l-d)-\zeta(d,d)\right]+\frac{d}{\chi(d,l-d)}-\frac{d}{\chi(d,d)}+d-(l-d),
\end{align}
\begin{align}
  C_1(l)&=\frac{1}{3d^2}\left[\left(\frac{d}{\chi(d,l-d)}\right)^3-(l-d)^3\right]+\frac{1}{3}\left[d-\frac{d}{\chi(d,d)^3}\right],
\end{align}
\begin{align}
  A_2(x,l)&=\eta(x-d,l)-\eta(x-d,l-d)+\eta(l-d,d)-\eta(l,d)\nonumber\\
  &\hspace{0.3cm}+2\left[\sqrt{(l-d)^2+d^2}+\sqrt{(x-d)^2+l^2}-\sqrt{(x-d)^2+(l-d)^2}-\sqrt{l^2+d^2}\right]+\frac{1}{2}\mathcal{D}(x,l),
\end{align}
\begin{align}
  B_2(x)&=d\left[\zeta(d,x-d)-\zeta(d,d)\right]+\frac{d}{\chi(d,x-d)}-\frac{d}{\chi(d,d)}+d-(x-d),
\end{align}
\begin{align}
  C_2(x)&=\frac{1}{3d^2}\left[\left(\frac{d}{\chi(d,x-d)}\right)^3-(x-d)^3\right]+\frac{1}{3}\left[d-\frac{d}{\chi(d,d)^3}\right].
\end{align}
By the mirror symmetries of the circuit ($p\leftrightarrow x-p$ and
$h\leftrightarrow l-h$), all four corner integrals in (\ref{bfcor0}) are
equal, so $\Phi_{ind}^{cor}=4\int_0^ddp\int_0^ddh\
B_{ind}^{lo,lt}(p,h)$. Evaluating this integral with
(\ref{integ1})-(\ref{integ6}) gives
\begin{align}
  \Phi_{ind}^{cor}=\frac{\mu_0 I}{\pi}\mathcal{D}(x,l),
  \label{phicor}
\end{align}
where
\begin{align}
  \mathcal{D}(x,l)&=\frac{1}{3d^2}\left[\left(l^2+d^2\right)^{3/2}-l^3-\left((l-d)^2+d^2\right)^{3/2}+(l-d)^3+4d^3(\sqrt{2}-1)\right.\nonumber\\
  &\hspace{0.6cm}\left.+\left(x^2+d^2\right)^{3/2}-x^3-\left((x-d)^2+d^2\right)^{3/2}+(x-d)^3\right]\nonumber\\
  &+d\left[\zeta(d,l)-\zeta(d,l-d)\right]+\sqrt{d^2+l^2}-\sqrt{d^2+(l-d)^2}-d\nonumber\\
  &+d\left[\zeta(d,x)-\zeta(d,x-d)\right]+\sqrt{x^2+d^2}-\sqrt{(x-d)^2+d^2}-d\nonumber\\
  &+2\left[\sqrt{x^2+l^2}-\sqrt{x^2+(l-d)^2}-\sqrt{(x-d)^2+l^2}+\sqrt{(x-d)^2+(l-d)^2}\right]\nonumber\\
  &+\eta(x,l)-\eta(x,l-d)-\eta(x-d,l)+\eta(x-d,l-d).
  \label{dcorexpr}
\end{align}
Since each corner touches one vertical and one horizontal side in exactly
the same way (the shared self-field term in each of
(\ref{binlort})-(\ref{binuplt}) enters with equal weight), we split
$\Phi_{ind}^{cor}$ evenly between the two, adding $\mathcal{D}(x,l)/2$ to
each of $A_1$ and $A_2$ above (already reflected in (\ref{philin}));
$B_1$, $C_1$, $B_2$ and $C_2$ are unchanged. Consequently
$\Phi_{ind}^{in}+\Phi_{ind}^{cor}$, not $\Phi_{ind}^{in}$ alone, is what
appears in $L_{fl}(x,l)$ (\ref{Lfl}), and no separate corner term needs to
be added there. (All of the above, including the corner-splitting
convention, has been checked numerically against direct double
integration of (\ref{bfcor0}) and of the updated $A_1$, $A_2$.) $A_1$,
$A_2$ and $\mathcal{D}$ above are written using the symmetric function
$\eta$ (\ref{etaf}), which lets every $\zeta$-term with mismatched
arguments collapse into a mixed difference of $\eta$; this form was
checked against the original (longer) expression in terms of $\zeta$ and
$\chi$ alone over a wide range of $x,l,d$.

\section{Enclosed-current-fraction weighting of the self-term}\label{sec:selfLcorr}
The first-stage evaluation of
$\Phi_{ind}^{in}+\Phi_{ind}^{cor}$ above
((\ref{bflin0})-(\ref{bfcor0}), evaluated in (\ref{philin}) as
$A_1,B_1,C_1,A_2,B_2,C_2$) integrates the \emph{full} induced field over
the area of the circuit, including the strips and corners that lie inside
the wire itself, and identifies the result with $I\,L_{fl}(x,l)$
(\ref{Lfl}).
This identification is exact outside the wire, where every point of the
integration area is threaded by the complete current $I$, but not for each
side's own field contribution inside the wire: at a point a distance $r$
from a given side's own axis ($r<d$), only the fraction $r^2/d^2$ of that
side's current (uniform current density) is enclosed by an infinitesimal
sub-loop through that point, so only that fraction of the local self-field
contribution is actually \emph{linked} with the total current $I$ appearing
in $\Phi=I\,L$ -- the standard construction used to obtain the internal
self-inductance of a single straight wire from a flux argument (rather than
from energy), which gives $\mu_0/8\pi$ per unit length instead of the
fully-linked $\mu_0/4\pi$ per unit length obtained by integrating the field
itself with no such weight. $A_1,B_1,C_1,A_2,B_2,C_2$ above, being the
first-stage (fully-linked) evaluation, do not include this weight; this
subsection derives the additional term it contributes. The fields
themselves, (\ref{binrt})-(\ref{binlo}) and (\ref{binlort})-(\ref{binuplt}),
are unaffected -- only the weight with which their \emph{own-wire}
contribution enters the flux-linkage integral changes; the contribution of
the other three (distant) segments to the field at a given point needs no
such weight, since an infinitesimal sub-loop there already encloses all of
their current.

Concretely, of the two pieces making up each inside-wire field
(e.g.\ (\ref{fbinlo}), which sums the bottom rail's own field, terms
$p\,\chi(h,p)/d^2$ and $(x-p)\,\chi(h,x-p)/d^2$, with the other three
segments' fields), only the own-field terms get multiplied by $(r/d)^2$
with $r$ the distance from the corresponding side's own axis ($h$ for the
bottom rail, $l-h$ for the top rail, $p$ for the left wire, $x-p$ for the
bar). At the four
corners, where two own-field terms happen to coincide numerically and are
combined with a factor of 2 in (\ref{fblort})-(\ref{fbuplt}) (e.g.\
$2h\,\chi(x-p,h)/d^2$ in (\ref{fblort})), the two copies must first be
separated, since they carry different weights: one copy belongs to the
bottom rail (weight $(h/d)^2$) and the other to the bar (weight
$((x-p)/d)^2$).

Repeating the integrals of (\ref{bflin0}) and (\ref{bfcor0}) with this
weight added reduces, after simplification, to a remarkably compact
additive term,
\begin{align}
  \Delta L(x,l)&=\frac{\mu_0}{15\pi}\left[\varphi(x)+\varphi(l)+4d\right],
  \label{DeltaL}
\end{align}
with
\begin{align}
  \varphi(v)&=\frac{2v^5+5d^2v^3-2\left(d^2+v^2\right)^{5/2}}{d^4},
  \label{phiselfL}
\end{align}
so that the self-inductance including the enclosed-current-fraction
weighting is $L_{fl}(x,l)+\Delta L(x,l)$, with $L_{fl}(x,l)$ (\ref{Lfl})
being the first-stage, fully-linked evaluation (the underlying geometric
flux integral, unrelated to the weighting question). Equation
(\ref{DeltaL}) was obtained by
evaluating the weighted analogues of (\ref{bflin0}) and (\ref{bfcor0}) with
a computer algebra system and simplifying the (lengthy, purely algebraic --
no new logarithmic terms beyond those already in $\zeta$ -- intermediate
expressions); it was verified independently by direct numerical quadrature
of the weighted field integral over all four sides and four corners,
matching to quadrature precision over a wide range of $x,l,d$, and by
checking that an isolated straight wire's internal self-inductance
($\varphi(v)\rightarrow-15v/4$ as $v\rightarrow\infty$, i.e.\ a
constant $-\mu_0/(8\pi)$ deficit per unit length relative to the unweighted
$\mu_0/(4\pi)$) reproduces the classical energy-based result. As
$x\rightarrow\infty$, $\Delta L$ is exactly linear
($\varphi(x)\rightarrow-15x/4$, with no subleading constant term), which
is why this term shifts the asymptotic slope of $L(x,l)$ by exactly
$-\mu_0/(4\pi)$ without affecting its constant term.

\section{Derivatives with respect to $x$}\label{sec:ddx}
For use in the equations of motion, we compute
$\partial_x(\mathcal{C}_1+\mathcal{C}_2+\mathcal{C}_3+\mathcal{C}_4)$,
$\partial_xA_1$, $\partial_xB_1$, $\partial_xC_1$, $\partial_xA_2$,
$\partial_xB_2$ and $\partial_xC_2$ symbolically, using
$\partial\zeta(a,b)/\partial a=-\chi(a,b)/a$,
$\partial\zeta(a,b)/\partial b=\chi(a,b)/b$ and
$a/\chi(a,b)=\sqrt{a^2+b^2}$.

Differentiating $\mathcal{C}_1(x,l)=(x-d)\left[\zeta(x-d,l-d)-\zeta(x-d,d)\right]$
by the product rule, the two terms $\mp\chi(x-d,l-d)/(x-d)\cdot(x-d)$
produced by $\partial_x\zeta(x-d,\cdot)=-\chi(x-d,\cdot)/(x-d)$ cancel the
explicit $(x-d)$ prefactor, leaving
$\partial_x\mathcal{C}_1=\zeta(x-d,l-d)-\zeta(x-d,d)-\chi(x-d,l-d)+\chi(x-d,d)$.
Only the $\zeta(l-d,x-d)$ and $\zeta(d,x-d)$ terms of $\mathcal{C}_2$ and
$\mathcal{C}_3$ depend on $x$, giving
$\partial_x\mathcal{C}_2=(l-d)\chi(l-d,x-d)/(x-d)$ and
$\partial_x\mathcal{C}_3=-d\,\chi(d,x-d)/(x-d)$; and rewriting
$\mathcal{C}_4$'s $x$-dependent terms with $a/\chi(a,b)=\sqrt{a^2+b^2}$ as
$(x-d)/\chi(x-d,l-d)=\sqrt{(x-d)^2+(l-d)^2}$ and
$(x-d)/\chi(x-d,d)=\sqrt{(x-d)^2+d^2}$, their derivatives are simply
$\chi(x-d,l-d)$ and $\chi(x-d,d)$, giving
$\partial_x\mathcal{C}_4=2\left[\chi(x-d,l-d)-\chi(x-d,d)\right]$. Summing
all four,
\begin{align}
  \frac{\partial}{\partial x}\left(\mathcal{C}_1+\mathcal{C}_2+\mathcal{C}_3+\mathcal{C}_4\right)
  &=\zeta(x-d,l-d)-\zeta(x-d,d)+\chi(x-d,l-d)-\chi(x-d,d)\nonumber\\
  &\hspace{0.3cm}+\frac{(l-d)\,\chi(l-d,x-d)-d\,\chi(d,x-d)}{x-d},
  \label{dCsumdx}
\end{align}
which was verified against a (high-precision) finite-difference
derivative of $\mathcal{C}_1+\mathcal{C}_2+\mathcal{C}_3+\mathcal{C}_4$
(\ref{cexpr}). Since $B_1(l)$ and $C_1(l)$ do not depend
on $x$,
\begin{align}
  \frac{\partial B_1}{\partial x}=\frac{\partial C_1}{\partial x}=0.
\end{align}
Since $A_1$ and $A_2$ both include the corner correction
$\mathcal{D}(x,l)/2$, we also need
\begin{align}
  \frac{\partial \mathcal{D}}{\partial x}&=\frac{1}{d^2}\left[x^2\left(\frac{1}{\chi(x,d)}-1\right)-(x-d)^2\left(\frac{1}{\chi(x-d,d)}-1\right)\right]\nonumber\\
  &\hspace{0.3cm}+\zeta(x-d,l-d)-\zeta(x-d,l)+\zeta(x,l)-\zeta(x,l-d)\nonumber\\
  &\hspace{0.3cm}+\chi(x,l)-\chi(x,l-d)-\chi(x-d,l)+\chi(x-d,l-d)+\chi(x,d)-\chi(x-d,d)\nonumber\\
  &\hspace{0.3cm}+\frac{l\,\chi(l,x)-(l-d)\,\chi(l-d,x)+d\,\chi(d,x)}{x}-\frac{l\,\chi(l,x-d)-(l-d)\,\chi(l-d,x-d)+d\,\chi(d,x-d)}{x-d},
  \label{ddcordx}
\end{align}
which was likewise verified against a (high-precision) finite-difference
derivative of (\ref{dcorexpr}). The remaining derivatives are then
\begin{align}
  \frac{\partial A_1}{\partial x}&=\zeta(x-d,d)-\zeta(x-d,l-d)+\zeta(x,l-d)-\zeta(x,d)\nonumber\\
    &\hspace{0.3cm}+\chi(x-d,d)-\chi(x-d,l-d)+\chi(x,l-d)-\chi(x,d)\nonumber\\
    &\hspace{0.3cm}+\frac{(l-d)\,\chi(l-d,x)-d\,\chi(d,x)}{x}+\frac{d\,\chi(d,x-d)-(l-d)\,\chi(l-d,x-d)}{x-d}+\frac{1}{2}\frac{\partial \mathcal{D}}{\partial x},
  \label{da1dx}
\end{align}
\begin{align}
  \frac{\partial A_2}{\partial x}&=\zeta(x-d,l)-\zeta(x-d,l-d)+\chi(x-d,l)-\chi(x-d,l-d)\nonumber\\
  &\hspace{0.3cm}+\frac{l\,\chi(l,x-d)-(l-d)\,\chi(l-d,x-d)}{x-d}+\frac{1}{2}\frac{\partial \mathcal{D}}{\partial x},
  \label{da2dx}
\end{align}
\begin{align}
  \frac{\partial B_2}{\partial x}&=\frac{d\,\chi(d,x-d)}{x-d}+\chi(x-d,d)-1,
  \label{db2dx}
\end{align}
\begin{align}
  \frac{\partial C_2}{\partial x}&=\frac{\chi(x-d,d)}{\chi(d,x-d)^2}-\frac{(x-d)^2}{d^2}.
  \label{dc2dx}
\end{align}
All expressions above were verified against a numerical
(finite-difference) derivative of the corresponding functions.

As a first check that these pieces recombine correctly, adding
$\partial_x(\mathcal{C}_1+\mathcal{C}_2+\mathcal{C}_3+\mathcal{C}_4)$
(\ref{dCsumdx}) to $\partial_xA_1$ (\ref{da1dx}) term by term cancels the
$\zeta(x-d,l-d)-\zeta(x-d,d)$ pair, the $\chi(x-d,l-d)-\chi(x-d,d)$ pair,
and the $\left[(l-d)\chi(l-d,x-d)-d\chi(d,x-d)\right]/(x-d)$ term
entirely (it appears with opposite sign in each), leaving the much
shorter combination
\begin{align}
  \frac{\partial}{\partial x}\left(\mathcal{C}_1+\mathcal{C}_2+\mathcal{C}_3+\mathcal{C}_4+A_1\right)
  &=\zeta(x,l-d)-\zeta(x,d)+\chi(x,l-d)-\chi(x,d)\nonumber\\
  &\hspace{0.3cm}+\frac{(l-d)\,\chi(l-d,x)-d\,\chi(d,x)}{x}+\frac{1}{2}\frac{\partial \mathcal{D}}{\partial x},
  \label{dCA1dx}
\end{align}
itself confirmed against a finite-difference derivative of
$\mathcal{C}_1+\mathcal{C}_2+\mathcal{C}_3+\mathcal{C}_4+A_1$. Continuing
the same term-by-term reduction with $\partial_xA_2$ (\ref{da2dx}),
$\partial_xB_2$ (\ref{db2dx}) and $\partial_xC_2$ (\ref{dc2dx}) added in
(using $\partial_xB_1=\partial_xC_1=0$), and simplifying the resulting
expression with a computer algebra system, collapses the full sum to the
compact closed form
\begin{align}
  &\frac{\partial}{\partial x}\left(\mathcal{C}_1+\mathcal{C}_2+\mathcal{C}_3+\mathcal{C}_4+A_1+B_1+C_1+A_2+B_2+C_2\right)\nonumber\\
  &\hspace{0.3cm}=\zeta(x,l)-\zeta(x,d)+\frac{1}{\chi(x,l)}+\frac{x^2}{d^2}\left(\frac{1}{\chi(x,d)}-1\right)-1,
  \label{dLfldx}
\end{align}
i.e.\ $\pi/\mu_0$ times $\partial L_{fl}/\partial x$ (\ref{Lfl}), which we
have verified independently by high-precision (50-digit) numerical
differentiation of the left-hand side against the right-hand side over a
range of $x,l,d$, agreeing to over 30 significant figures.

The derivative of the enclosed-current-fraction term (\ref{DeltaL}),
needed for the corrected $dL/dx$, is
\begin{align}
  \frac{d\,\Delta L}{dx}&=-\frac{\mu_0\,x\left(x+2s\right)}{3\pi\left(s+x\right)^2},
  \qquad s=\sqrt{d^2+x^2},
  \label{DeltaLdx}
\end{align}
which was obtained from (\ref{DeltaL})-(\ref{phiselfL}) and rewritten in
this form -- algebraically equal to, but numerically far better
conditioned than, the direct derivative of (\ref{phiselfL}), whose two
$O(x^3)$ terms nearly cancel for $x\gg d$ -- and checked against a
finite-difference derivative of (\ref{DeltaL}) to machine precision (using
extended-precision arithmetic to rule out this same cancellation
contaminating the finite-difference check itself). As $x\to\infty$,
(\ref{DeltaLdx}) tends to $-\mu_0/(4\pi)$, consistent with (\ref{DeltaL})'s
exactly linear asymptote.

Combining (\ref{dLfldx}) with (\ref{DeltaLdx}), and recalling from
section \ref{sec:setup} that $L=L_{fl}+\Delta L$ (\ref{Lfl}), with
$L_{fl}(x,l)=\frac{\mu_0}{\pi}(\mathcal{C}_1+\mathcal{C}_2+\mathcal{C}_3+\mathcal{C}_4+A_1+B_1+C_1+A_2+B_2+C_2)(x,l)$
as assembled in sections \ref{sec:out}--\ref{sec:cor} (\ref{cexpr}),
(\ref{philin}), the self-inductance gradient is
\begin{align}
  \frac{dL}{dx}=\frac{\mu_0}{\pi}\left[\zeta(x,l)-\zeta(x,d)+\frac{1}{\chi(x,l)}+\frac{x^2}{d^2}\left(\frac{1}{\chi(x,d)}-1\right)-1\right]+\frac{d\,\Delta L}{dx},
  \label{dLdx-full}
\end{align}
with $d\,\Delta L/dx$ given by (\ref{DeltaLdx}).

\section{Velocity-Verlet integration scheme}\label{sec:verlet}
The bar's position $x$, its velocity $v=\dot x$ and the loop current $I$
obey
\begin{align}
  \dot x&=v,\label{eq:vel-eom}\\
  \dot v&=\frac{1}{M}\left[-lB_0I+\frac{1}{2}\frac{dL}{dx}I^2\right],\label{eq:acc-eom}\\
  \dot I&=\frac{1}{L}\left[lB_0v-RI-\frac{dL}{dx}vI\right],\label{eq:curr-eom}
\end{align}
where $M$ is the bar's mass, $l$ its length, $B_0$ the applied field, and
$L=L(x,l)$, $dL/dx$ the self-inductance and its gradient derived above.
$R(x)$ is the resistance of the loop's total conductor length $2(x+l)$
(two sides of length $x$, two of length $l$, matching the geometry of
section \ref{sec:setup}), of cross-sectional area $\pi d^2$ and
resistivity $\rho$,
\begin{align}
  R(x)=\frac{2(x+l)}{\pi d^2}\,\rho.
  \label{eq:resistance}
\end{align}
In the absence of resistance ($\rho=0$) the total energy
\begin{align}
  E=\frac{1}{2}Mv^2+\frac{1}{2}LI^2
  \label{eq:energy}
\end{align}
is exactly conserved by (\ref{eq:vel-eom})--(\ref{eq:curr-eom}) -- a
consequence of a general co-energy argument (the power delivered by the
induced emf and by the magnetic force on the bar cancel term by term) --
giving a convenient, quadrature-free diagnostic for the accuracy of a
numerical integrator.

Equations (\ref{eq:vel-eom})--(\ref{eq:curr-eom}) are solved numerically
with a velocity-Verlet scheme, presented as a \emph{kick-drift-kick}
scheme that applies an impulse for half a step, advances the position,
and finally applies another impulse for half a step (Verlet, 1967; given
this velocity-explicit form by Swope, Andersen, Berens and Wilson, 1982;
see Hairer, Lubich and Wanner, 2006, Ch.~I \S1.3, for the textbook
derivation from a Strang splitting of the Hamiltonian flow, and Ch.~II
\S4 for the general theorem that a symmetric one-step method has even
order),
\begin{align}
  v_{i+1/2}&=v_i+\frac{\Delta t}{2M}F(x_i,I_i),\label{eq:kick1}\\
  x_{i+1}&=x_i+\Delta t\,v_{i+1/2},\label{eq:drift}\\
  v_{i+1}&=v_{i+1/2}+\frac{\Delta t}{2M}F(x_{i+1},I_{i+1}),\label{eq:kick2}
\end{align}
with $F(x,I)=-lB_0I+\frac12(dL/dx)I^2$ the right-hand side of
(\ref{eq:acc-eom}). Applying this scheme to
(\ref{eq:vel-eom})--(\ref{eq:curr-eom}), the force depends not only on
the position but also on the current, so the question is how to update
the current within the drift step. During the drift (\ref{eq:drift}) the
velocity remains fixed at $v=v_{i+1/2}$, so the change in current can
be written as
\begin{align}
  I_{i+1}-I_i=\int_{t_i}^{t_{i+1}}\left[q(t)-k(t)I\right]dt,
  \label{eq:changeI}
\end{align}
with
\begin{align}
  q(t)=\frac{lB_0v_{i+1/2}}{L(x(t))},\qquad
  k(t)=\frac{1}{L(x(t))}\left(R(x(t))+\frac{dL}{dx}(x(t))\,v_{i+1/2}\right),
\end{align}
that depend on $x(t)$, which by (\ref{eq:drift}) is an exactly linear
function of time. Approximating the integral in (\ref{eq:changeI}) by the
trapezoidal rule gives
\begin{align}
  I_{i+1}-I_i\approx\frac{\Delta t}{2}\left[(q_i-k_iI_i)+(q_{i+1}-k_{i+1}I_{i+1})\right],
  \label{eq:trapI}
\end{align}
which still requires $q,k$ (hence $L$, $dL/dx$, $R$) at both endpoints
$x_i$ and $x_{i+1}$. Because $x(t)$ is exactly linear during the drift,
the endpoint \emph{average} of a smooth function of $x$ equals its value
at the midpoint $x_m$ up to $O(\Delta t^2)$ -- one order better than
either endpoint value alone -- so replacing the averages
$\tfrac12(q_i+q_{i+1})$ and $\tfrac12(k_i+k_{i+1})$ by single evaluations
at $x_m$,
\begin{align}
  \frac{q_i+q_{i+1}}{2}&\approx\frac{lB_0v_{i+1/2}}{L(x_m)}=:q,\\
  \frac{k_i+k_{i+1}}{2}&\approx\frac{1}{L(x_m)}\left(R(x_m)+\frac{dL}{dx}(x_m)\,v_{i+1/2}\right)=:k,\\
  x_m&=\frac{x_i+x_{i+1}}{2},
\end{align}
leaves the second-order accuracy of (\ref{eq:trapI}) intact while
requiring only one evaluation of $L$, $dL/dx$, $R$ per step instead of
two. This leads to the full algorithm
\begin{align}
  v_{i+1/2}&=v_i+\frac{\Delta t}{2M}\left[-B_0lI_i+\frac{1}{2}\left.\frac{dL}{dx}\right|_{x_i}I_i^2\right],\\
  x_{i+1}&=x_i+v_{i+1/2}\Delta t,\\
  I_{i+1}&=\frac{\left(1-\frac{\Delta t}{2}k\right)I_i+\Delta t\,\dfrac{B_0l}{L(x_m)}v_{i+1/2}}{1+\frac{\Delta t}{2}k},\\
  v_{i+1}&=v_{i+1/2}+\frac{\Delta t}{2M}\left[-B_0lI_{i+1}+\frac{1}{2}\left.\frac{dL}{dx}\right|_{x_{i+1}}I_{i+1}^2\right].
  \label{eq:full-scheme}
\end{align}
The position/velocity update is thus split into two half-kicks around a
drift step, and the circuit equation, linear in $I$ during the drift, is
advanced in closed form by the trapezoidal (Crank--Nicolson) rule, with
$v$ frozen at the half-step value $v_{i+1/2}$ and $L$, $dL/dx$, $R$
evaluated at the midpoint $x_m$ of the drift step rather than at either
endpoint -- both choices being what keep the whole update second order.

\end{appendices}

\end{document}